\documentclass[aps,prb,twocolumn,superscriptaddress]{revtex4-2}

\usepackage{amsmath}
\usepackage{graphicx}
\usepackage{epstopdf}
\usepackage{natbib}
\usepackage{array}
\usepackage{dcolumn}
\usepackage{bm}
\usepackage{appendix}
\usepackage{soul}  
\usepackage{multirow}
\usepackage{braket}
\usepackage{bbold}
\usepackage{hyperref}
\usepackage[usenames,dvipsnames]{xcolor}

\newcommand{\nv}{NV$^-$} 
\newcommand{\cbcn}{C$_{\text{B}}$C$_{\text{N}}$}
\newcommand{\feal}{$\text{Fe}_{\text{Al}}$}

\begin{document}
\title{Quantum embedding methods for correlated excited states of point defects: \\Case studies and challenges}
\author{Lukas Muechler}
\affiliation{Center for Computational Quantum Physics, Flatiron Institute, 162 5$^{th}$ Avenue, New York, NY 10010}
\author{Danis I. Badrtdinov}
\affiliation{Theoretical Physics and Applied Mathematics Department,
Ural Federal University, 620002 Yekaterinburg, Russia}
\affiliation{Center for Computational Quantum Physics, Flatiron Institute, 162 5$^{th}$ Avenue, New York, NY 10010}
\author{Alexander Hampel}
\affiliation{Center for Computational Quantum Physics, Flatiron Institute, 162 5$^{th}$ Avenue, New York, NY 10010}
\author{Jennifer Cano}
\affiliation{Center for Computational Quantum Physics, Flatiron Institute, 162 5$^{th}$ Avenue, New York, NY 10010}
\affiliation{Department of Physics and Astronomy, Stony Brook University, Stony Brook, New York 11794-3800, USA}
\author{Malte R\"osner}
\affiliation{Radboud University, Institute for Molecules and Materials, Heijendaalseweg 135, 6525 AJ Nijmegen, Netherlands}
\author{Cyrus E. Dreyer}
\affiliation{Center for Computational Quantum Physics, Flatiron Institute, 162 5$^{th}$ Avenue, New York, NY 10010}
\affiliation{Department of Physics and Astronomy, Stony Brook University, Stony Brook, New York 11794-3800, USA}
\date{\today}

\begin{abstract}
A quantitative description of the excited electronic states of point defects and impurities is crucial for understanding materials properties, and possible applications of defects in quantum technologies. This is a considerable challenge for computational methods, since Kohn-Sham density-functional theory (DFT) is inherently a ground state theory, while higher-level methods are often too computationally expensive for defect systems. Recently, embedding approaches have been applied that treat defect states with many-body methods, while using DFT to describe the bulk host material. 
We implement such an embedding method, based on Wannierization of defect orbitals and the constrained random-phase approximation approach, and perform systematic characterization of the method for three distinct systems with current technological relevance: a carbon dimer replacing a B and N pair in bulk hexagonal BN (\cbcn{}), the negatively charged nitrogen-vacancy center in diamond (\nv{}), and an Fe impurity on the Al site in wurtzite AlN (\feal{}). For \cbcn{} we show that the embedding approach gives many-body states in agreement with analytical results on the Hubbard dimer model, which allows us to elucidate the effects of  the DFT functional and double-counting correction. For the \nv{} center, our method demonstrates good quantitative agreement with experiments for the zero-phonon line of the triplet-triplet transition. Finally, we illustrate challenges associated with this method for determining the energies and orderings of the complex spin multiplets in \feal{}.
\end{abstract}

\maketitle

\section{Introduction}  \label{sec:intro}
Point defects, such as vacancies, interstitial atoms, antisites, and atomic impuritites, are ubiquitous in all materials. Even when present in minute concentrations, they can profoundly alter material and device properties. Defects are often detrimental to device performance; for example, so-called Shockley-Read-Hall (SRH)\cite{Shockley1952,Hall1952} defect-mediated recombination of electrons and holes in semiconductors is a key efficiency-limiting process in solar cells and light-emitting diodes. However, more recently, defects have emerged as robust and manipulatable quantum systems for the next generation of quantum technologies, e.g., spin qubits for quantum computing \cite{Weber2010,Kane1998,Pla2012,Wu2019}, single-photon emitters (SPEs) for quantum communication \cite{Aharonovich2011_RPP,Aharonovich2011}, and nanoprobes for quantum metrology \cite{Schirhagl2014}.

In both contexts, the properties of the \emph{electronic excited states} of the defect play a key role. 
For quantum applications, manipulation of the spin-qubit state for computing is often carried out via optical excitation, and relies on specific nonradiative transitions from the excited state (i.e., intersystem crossings \cite{Doherty2013,Thiering2018}). Also, whether or not a defect will be appropriate as a SPE depends on the electron-phonon coupling of the defect in its excited state \cite{Aharonovich2011_RPP}. Finally, nanometrology with defects often relies on the dipole moment or magnetic properties of the excited states \cite{Schirhagl2014}. Point-defect excited states also play an important role when considering their detrimental effect on the host material, for example resulting in additional channels for SRH in wide-band-gap insulators \cite{Alkauskas2016,Wickramaratne2016}. 

Thus, a quantitative theoretical understanding of the electronic excited states of defects is crucial. However, describing defect excited states from first-principles is a significant challenge. Kohn-Sham density functional theory (KS-DFT), which is the workhorse for determining
defect properties \cite{Dreyer2018,Freysoldt2014}, is a ground-state theory, and 
the calculated eigenvalues do not correspond to the quasiparticle addition/removal energies \cite{Perdew1981,MorisSanchez2008}. As is the case for atoms or molecules, the excited states may correspond
to multiplets that cannot be described by a single-Slater-determinant
theory like KS-DFT \cite{vonBarth1979,Lischner2012}. 

This motivates the use of higher-level many-body methods to treat defect excited states; however,
defects are a challenging application for such methods,  due to computational expense. Specifically, in order to model an isolated defect, large ``supercells'' are necessary to separate defects from their periodic images; if open boundary conditions are used, then large clusters are required to converge to a bulk-like environment for the defect.


The computational problem of treating a small ``active space'' of correlated defect states within a relatively weakly correlated bulk (i.e., supercell or cluster) is thus ideal for a quantum embedding approach \cite{Qiming2016,Jones2020-ea}. Such approaches have enjoyed extensive success in quantum chemistry \cite{Pascual1995-dt,Llusar1996-lx,Kluner2002-um,Munoz_Ramo2007,Gomes2008-mt,Swerts2008-rm,Swerts2008-rq,Huang2008-hi,Pascual2009-le,Gomes2012-mb,Goodpaster2014-tw,Nguyen_Lan2016-tk,Dvorak2019-gf,Jones2020-ea,Cui2020-zr,Sriluckshmy2021-ou,He2022-lq} and solid-state physics \cite{Kotliar2006,Haule2018,Held2007,Qiming2016} for treating strongly correlated materials and molecules. Recently, their popularity for treating defects \cite{Galli2021,Bockstedte2018,Ma2020,Ma2020_2,Barcza2021,Pfaffle2021} and other inhomogeneous systems \cite{Gardonio2013,Fenjie2015,Eskridge2019,Virgus2012,Virgus2014,Zhang2019-kd,Schafer2021-xi,Schafer2021-tn,Schafer2021-tn,Lau2021} has increased rapidly.

For the case of defects in semiconductors and insulators, the methodology demonstrated first by Bockstedte \textit{et al.}~\cite{Bockstedte2018}, and then implemented and developed by other groups \cite{Galli2021,Ma2020,Ma2020_2,Barcza2021,Pfaffle2021}, is particularly promising. This approach combines the state-of-the-art and highly successful methods for DFT calculations of defects in semiconductors \cite{Freysoldt2014} with downfolding and embedding approaches tailored to solid-state systems \cite{Kotliar2006,Haule2018,Held2007,Aryasetiawan2006}.
Specifically, the basis for this embedding approach \cite{Bockstedte2018,Galli2021,Ma2020,Ma2020_2} is a DFT calculations of the defect in a periodic supercell, which avoids the challenges associated with finite clusters sometimes used in quantum chemistry implementations, including quantum confinement effects and interactions between defect wave functions and the cluster surface (see Ref.~\onlinecite{Freysoldt2014} for discussion of supercells versus clusters). The choice of DFT is also found to be a much better starting point for such systems compared to Hartree-Fock, which is widely used in quantum chemistry implementations. Also, the active space is chosen to be minimal, and a screened Coulomb interaction is used in that space;  this is often the preferred approach in solid-state embedding \cite{Kotliar2006,Haule2018,Held2007,Aryasetiawan2006}, as opposed to increasing the size of the active space with a bare Coulomb interaction towards convergence (see, e.g., Ref.~\onlinecite{Zhu2021}). Owing to the small size of the Hilbert space, our method will serve as the basis for developing simplified effective models that capture qualitative and quantitative aspects of the system.

One of the key challenges of these methods is developing a quantitatively accurate \textit{ab-initio}  procedure for downfolding onto the active space. The details by which the DFT calculation in the bulk is combined with the MB calculation in the active subspace are important for accurate final observables. These details include: (i) the choice of the initial electronic configuration on which to base the embedding methodology; (ii) the procedure for isolating the correlated orbitals from the bulk; (iii) the approach for obtaining the effective Coulomb interaction in the subspace \cite{Aryasetiawan2004,Aryasetiawan2006, Galli2021}; and (iv) the approach to avoid ``double-counting'' errors of the Coulomb interaction as a result of the  DFT starting point \cite{Karolak2010,Haule2015}.

In this work we will explore these issues with the goal of developing quantum embedding techniques~\cite{Bockstedte2018,Galli2021} for quantitative prediction of defect properties in a variety of systems. 
To this end, we perform calculations on three diverse test-case defects, with a focus on systematic characterization of the methodological details (i)-(iv) above.
The first is a carbon dimer replacing a boron and nitrogen atom (\cbcn{}) in bulk hexagonal BN, whose simple electronic structure will allow comparison of the results to model calculations. The second is the \nv{} center in diamond, which is the prototypical correlated defect, and will serve as a benchmark against experiment and other computational techniques. The third is an iron atom replacing aluminum in AlN (\feal{}), which will serve as a stringent test of the methodology on a defect where correlations play a key qualitative role.

The rest of the paper is organized as follows: in Sec.~\ref{sec:defects} we introduce the case-study defects, the detailed motivation for choosing them, and their electronic structure; Section~\ref{sec:method} provides a brief outline of the general embedding approach; we analyze in detail the aspects of the methodology mentioned above in the context of our test-case defects in Sec.~\ref{sec:lessons}; in Sec.~\ref{sec:disc} we discuss some additional aspects of the methodology, including how to quantify the correlated nature of the MB states and how our method can be used to generate simplified models; we conclude the paper in Sec.~\ref{sec:conclusions}.

\section{Case-study defects \label{sec:defects}}

In this section, we will briefly introduce the case-study defects, their electronic structure, and the motivation for why they were chosen for this study. Computational parameters for each defect can be found in the supplemental materials (SM) \cite{SM} Sec.~S1.

To discuss the many-body (MB) states of the defects, we will adopt the following notation:
\begin{equation}
\vert \overline{\phi_1\phi_2...\phi_{N_{\text{orb}}}};\phi_1\phi_2...\phi_{N_{\text{orb}}}\rangle= \prod_{i} c_{i\downarrow}^\dagger \prod_{j} c_{j\uparrow}^\dagger  \vert 0\rangle,
\end{equation}
where  $\vert 0\rangle$ is the vacuum, $N_\text{orb}$ is the number of orbitals, $\phi_i$ labels the defect spin-orbital (spin-orbit interaction will not be treated in this work), and is either 0 or 1 indicating the occupancy of the basis spin-orbital state, the overbar indicates spin down,
and $c_{i\downarrow}^\dagger$ ($c_{j\uparrow}^\dagger$) only appears on the right-hand-side when $\overline{\phi_i}=1$ ($\phi_j=1$). 

\subsection{\cbcn{} in BN: A simple model} \label{sec:cbcn}

\cbcn{} in hexagonal BN has attracted significant recent attention, as it was proposed as the origin of the 4.1 eV zero-phonon line \cite{Era1981,Museur2008,Du2015} (ZPL, see Sec.~\ref{sec:zpl}) single-photon emitter observed in BN based on the energetics of emission \cite{Mackoit2019}, and calculations of photoluminescence lineshapes  \cite{Linderalv2021,Jara2021}. 
For our purposes, \cbcn{} was chosen as it has a particularly simple electronic structure that can be compared to analytical calculations.

\begin{figure}
   \includegraphics[width=\columnwidth]{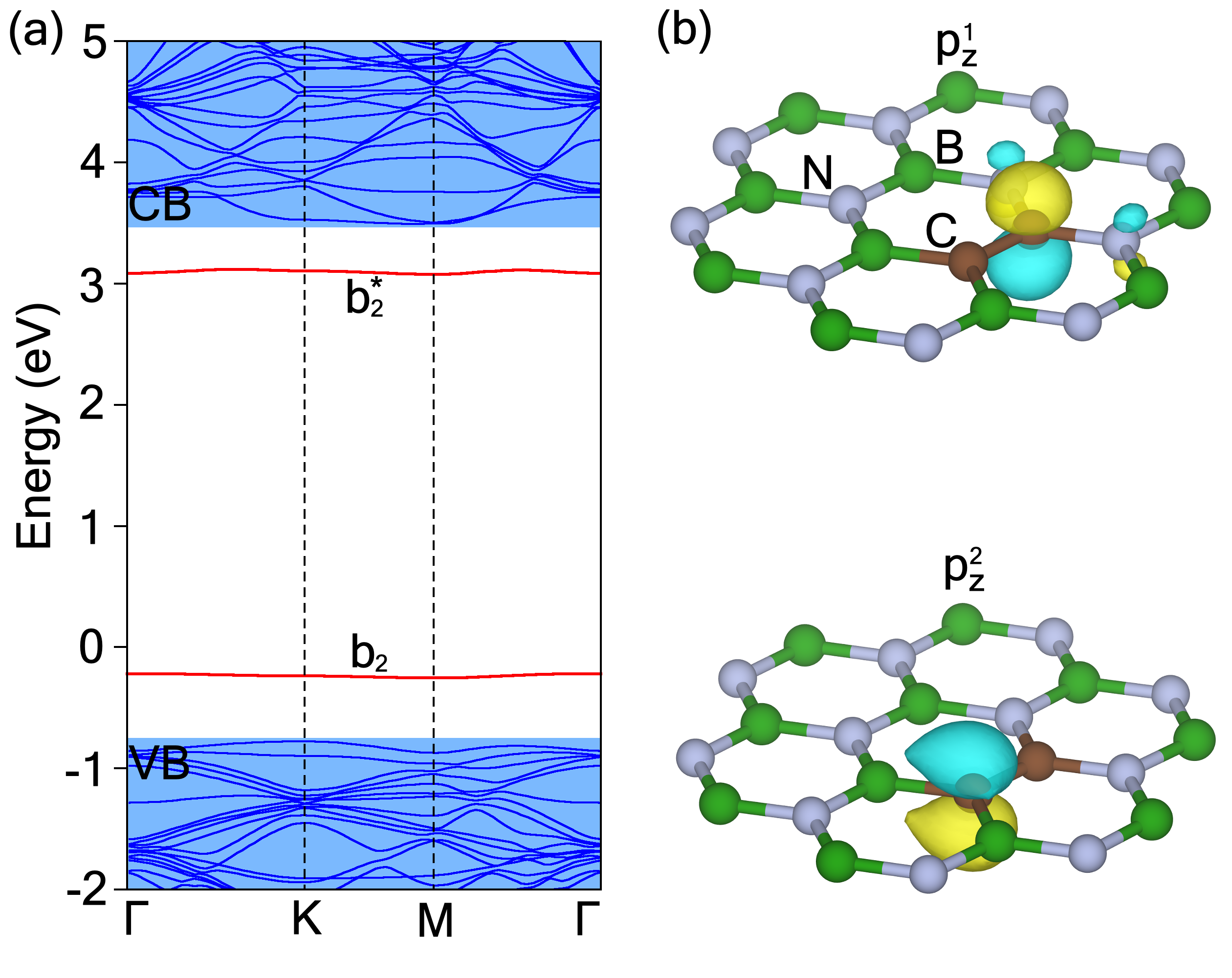}
\caption{\label{fig:cbcn_bs} (a) DFT Band structure for \cbcn{} in BN calculated with the PBE functional. Defect states are highlighted with red color and denoted with symmetry labels; bulk bands are blue. (b) The carbon $p_z$-character Wannier functions used to define the active space. Calculations are performed on bulk BN; a single 2D layer is shown here for clarity.}
\end{figure}


As a host material, we will consider bulk $P6_3/mmc$, i.e., three-dimensional layered BN. Replacing a nearest neighbor B and N with a neutral C dimer results in two defect states within the band gap of BN which are of bonding and anti-bonding character resulting from corresponding combinations of C $p_z$ orbitals~\cite{Mackoit2019}, c.f., Fig.~\ref{fig:cbcn_bs}(a). We label these states $b_2$ and $b_2^*$, respectively, using the irreducible representation (irrep) in the $C_{2v}$ point group of the defect and use them as the active space of the defect. The defect MB states can correspondingly be expressed in two equivalent bases: 
$\vert \overline{b_2} \overline{b^*_2}; b_2 b^*_2\rangle$, which we call the band basis, since it corresponds to the basis of Kohn-Sham bands; and $\vert \overline{p^2_z} \overline{p^1_z}; p^2_z p^1_z\rangle$, which we refer to as the orbital basis, since it will be the localized Wannier basis we use [see Fig.~\ref{fig:cbcn_bs}(b) and Sec.~\ref{sec:tijwan}]. As discussed in Sec.~\ref{sec:models}, calculations in these bases are equivalent, but the physical interpretation as well as the development of simplified models may be more transparent for a given choice.

Based on comparison with the simple model of a Hubbard dimer (see SM \cite{SM} Sec.~ S2 A 1), we expect six MB states: a ground-state $\vert\text{GS}\rangle$ spin singlet, a spin triplet manifold $\vert\text{T}\rangle$, and two additional spin singlets, $\vert\text{D}\rangle$ and $\vert\text{DS}\rangle$. More details of these states are given in the SM \cite{SM} Sec.~S2 A.

\subsection{\nv{} in diamond: An experimental benchmark} \label{sec:nv}

\begin{figure}
   \includegraphics[width=\columnwidth]{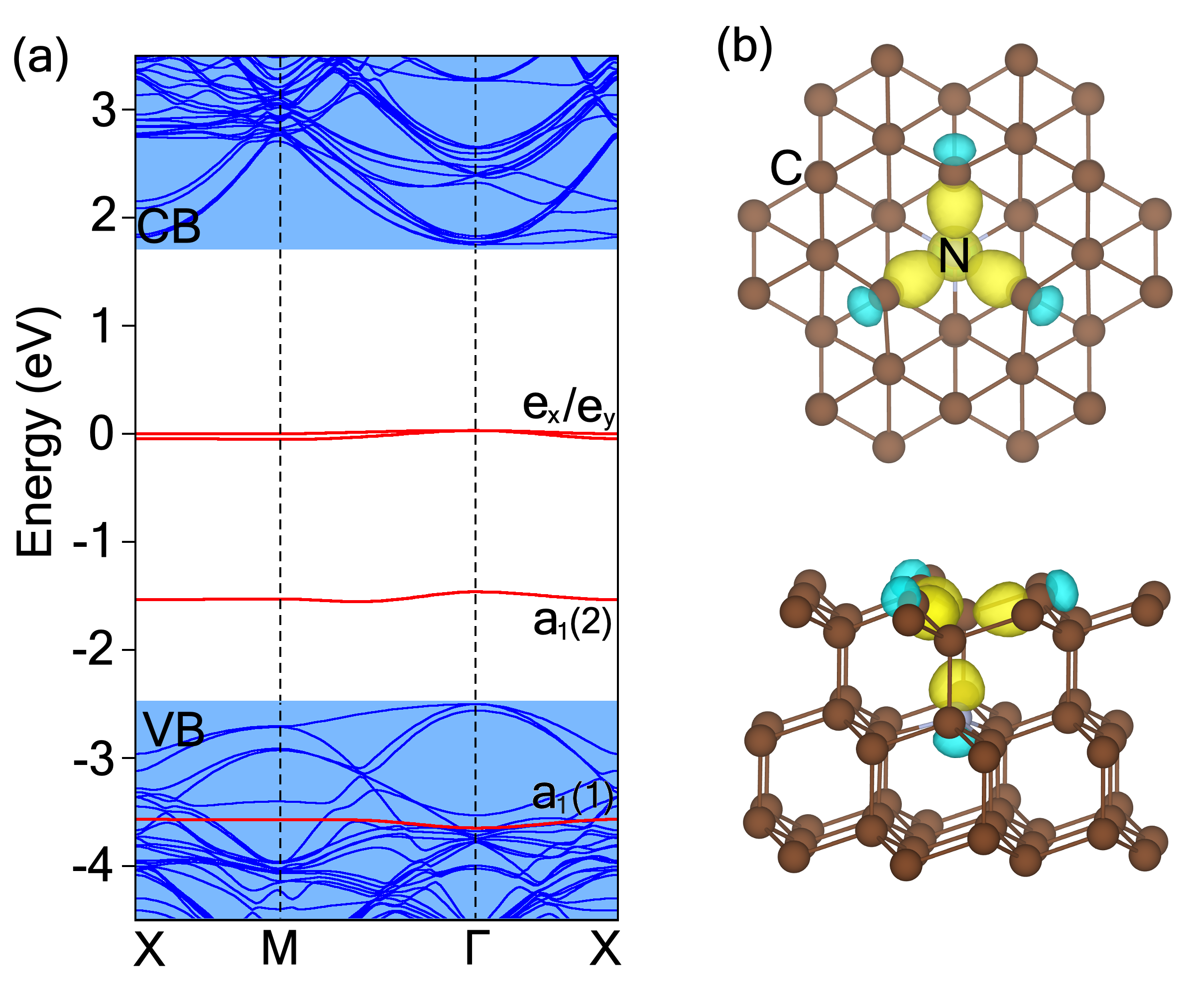}
\caption{\label{fig:NV_elec_struct} (a) Band structure for \nv{} center in diamond calculated with PBE. Defect states are highlighted with red color and denoted with symmetry labels, the bulk bands in blue. (b) The maximally-localized Wannier functions plotted in top and side view are shown for the $sp^3$ dangling bonds of N  and C atoms, which were used to define the active space.}
\end{figure}

The negatively-charged nitrogen-vacancy center in diamond (\nv{})
is the prototypical deep defect for quantum technologies \cite{Doherty2013}. The motivation to choose it for this study is that it is extremely well-characterized experimentally, and with a variety of theoretical methods. Thus it has become a standard for experimental verification of theories for correlated defect states (for reviews see, e.g., Refs.~\onlinecite{Doherty2013,Gali2019,Acosta2013,Aharonovich2011,Aharonovich2011_RPP,Schirhagl2014}).

The \nv{} center consists of a C vacancy in diamond with a N substituting a nearest-neighbor C atom. The defect states consist of dangling $sp^3$ bonds from the carbons and nitrogen atoms around the vacancy \cite{Doherty2011,Maze2011}. Linear combinations of these dangling bonds results in the single-particle defect states shown in Fig.~\ref{fig:NV_elec_struct}(a). Three states are in the bulk band gap, labeled by their irreps of the $C_{3v}$ point group  as $a_1(2)$ (which is doubly occupied) and $e$ (which is two-fold degenerate, also occupied by two electrons). There is another doubly-occupied $a_1(1)$ state resonant with the valence band (VB). 

As with \cbcn{}, we will consider two bases for constructing MB states. In this case, the band basis will be labelled by the irreps of the Kohn-Sham states $\vert \overline{e_x e_y a_1(2) a_1(1)} ; e_xe_ya_1(2)a_1(1)\rangle$, while the orbital basis will consists of the $sp^3$ dangling bond of the carbons next to the vacancy (labelled 1-3) and the N: $\vert \overline{sp^3_{\text{C}_1}sp^3_{\text{C}_2}sp^3_{\text{C}_3}sp^3_{\text{N}}} ; sp^3_{\text{C}_1}sp^3_{\text{C}_2}sp^3_{\text{C}_3}sp^3_{\text{N}}\rangle$ [see Fig.~\ref{fig:NV_elec_struct}(b)].

The MB states of \nv{}, as determined from symmetry considerations and orbital models \cite{Maze2011}, experiment \cite{Doherty2011}, and previous calculations \cite{Bockstedte2018,Ma2020,Bhandari2021}, consist of: a ground state triplet $^3A_2$, represented by aligned spins on $e_x$ and $e_y$; an excited state singlets $^1E_1$ from flipping one of the spins in the $e$ states; and an excited state singlet $^1A_1$ and triplet $^3E$ resulting from exciting an electron from $a_1(2)$ to the $e$ manifold. Only the fully symmetric atomic configuration of $^3E$ is considered in this work, i.e., we neglect the small energy reduction from the Jahn-Teller splitting of that degenerate state \cite{Thiering2017}. See SM \cite{SM} Sec.~S3 for details on how we calculate the symmetry of the MB states.

\subsection{\feal{} in AlN: A challenging correlated state }\label{sec:feal}

\begin{figure}
   \includegraphics[width=0.50\textwidth]{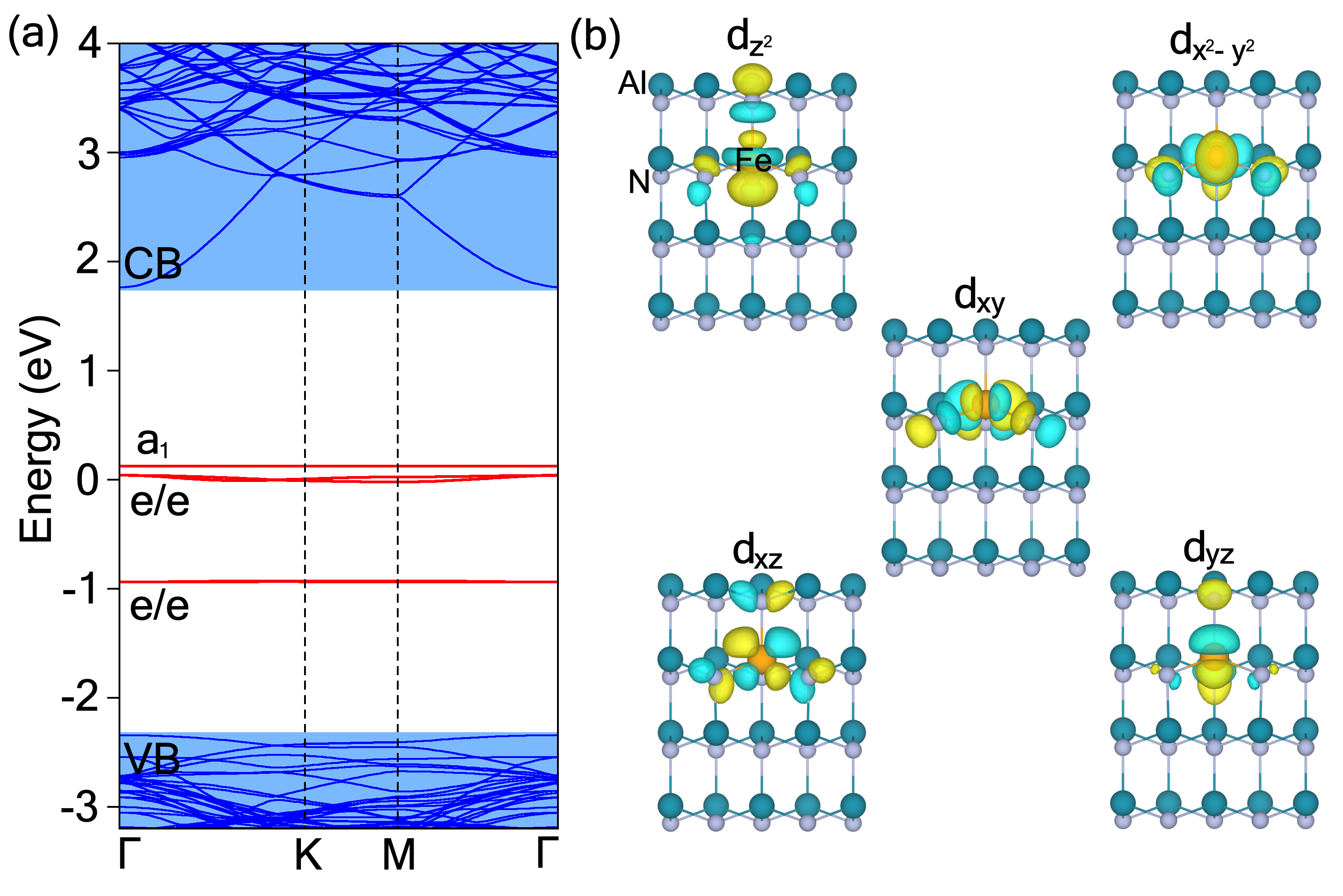}
\caption{\label{fig:feal_el_struct} (a) Band structure for \feal{} in wurtzite AlN calculated with the PBE functional. Defect states are highlighted with red color and denoted with symmetry labels, the bulk bands in blue. (b) The  Wannier functions for $3d$ orbital, which were used to define the active space.}
\end{figure}

The final defect we will consider in this study is an Fe atom substituted on an Al site (\feal{}) in wurtzite AlN (Fig.~\ref{fig:feal_el_struct}). In general, transition-metal (TM) impurities in semiconductors and insulators have
been widely studied as both detrimental and functional defects; e.g.,
Cu and Au are notorious deep traps/SRH recombination centers in Si
\cite{Weber1983}, while Cr in Al$_2$O$_3$ is the famous color center
responsible for the red emission of ruby \cite{Maiman1960}. Often,
transition-metal impurities have an open $d$-shell, and thus a rich
structure of multiplet excited states, which may
be interesting for spin qubits.

The reason that we chose \feal{} for this study is that it represents a significantly more complex electronic structure than both \cbcn{} and \nv{}. As we will show, rather than correlations renormalizing the single-particle picture, they qualitatively change the nature of the states. At the same time, 
the relevant defect states reside inside of the band gap. This removes effects of disentanglement, which we will not discuss in detail in this work.

We will focus on the neutral charge state of \feal{}, which
was found to be the lowest energy for Fermi levels near the mid gap of AlN \cite{Wickramaratne2019}. 
In this case, \feal{} exists in the $3+$ oxidation state, and thus has five electrons in the Fe $3d$ states. The point symmetry of \feal{} in wurtzite AlN is $C_{3v}$, though it is often assumed that the symmetry breaking from $T_d$ (as would be the case for \feal{} in \emph{zincblende} AlN) is small. The $T_d$ crystal field splits the Fe $d$ orbitals into a lower-energy, doubly degenerate $e$ manifold, and a higher energy $t_2$ manifold. The lower-symmetry crystal field of wurtzite AlN splits the $t_2$ states into $e$ and $a_1$ states [Fig.~\ref{fig:feal_el_struct}(a)].
Without spin polarization, these states are in the band gap near the VB, as shown in Fig.~\ref{fig:feal_el_struct} (note that including spin polarization significantly changes the nature of the states \cite{Wickramaratne2019}, as discussed in Sec.~\ref{sec:initial}). Thus the basis for studying \feal{} is simply these $3d$ states:  $\vert \overline{d_{z^2}}\overline{d_{xz}}\overline{d_{yz}}\overline{d_{x^2-y^2}}\overline{d_{xy}};d_{z^2}d_{xz}d_{yz}d_{x^2-y^2}d_{xy}\rangle$. 


For the other test case defects, we knew exactly the nature of the MB states that we should find in our embedding calculations. However \feal{} is much less explored theoretically and experimentally, and thus serves as a test of the predictive power of our method. Still, we can use ligand field theory for a general $d^5$ ion in a $C_{3v}$ crystal field as the basis of our expectations for the MB states \cite{Sugano1970,Malguth2006,Malguth2008,Neuschl2015,Wickramaratne2019}. Starting with $T_d$, the nature of the ground and excited states is determined by the magnitude of the crystal-field splitting (CFS). For relatively small CFS, the ground state is a 6-fold degenerate spin $5/2$ state with $A_1$ orbital symmetry (denoted $^6A_1$) that originates from the high-spin $^6S$ ground state of the free atom. The low-lying excited states are split from the spin $3/2$ states of the free atom $^4G$: $^4T_1$, $^4T_2$, $^4E$, and $^4A_1$, for $T_d$ \cite{Malguth2006,Malguth2008,Sugano1970}; in $C_{3v}$, the $^4T_1$ ($^4T_2$) irreps split into a $^4E$ and $^4A_2$ ($^4A_1$). 
For large CFS, the ground state becomes the three-fold degenerate, spin $1/2$ $^2T_2$ state, and the low energy excited states may include various other spin $1/2$ states originating from the $^2I$ manifold of the free atom in addition to the spin $3/2$ states.

It has been experimentally demonstrated \cite{Malguth2006,Malguth2008,Neuschl2015} that Fe$_{\text{Ga}}$ in GaN in the neutral charge state has a high-spin $S=5/2$ ground state, and all theoretical \cite{Wickramaratne2019,Zakrzewski2016} and experimental \cite{Baur1994,Soltamov2010,Masenda2016} work on \feal{} in AlN indicate that it should be the same. As we will see in Sec.~\ref{sec:DC}, determining the ground state to be high-spin versus low-spin is a sensitive quantitative test of our methodology.

\section{General approach} \label{sec:method}

Now that we have introduced our case-study defects, we  will give a brief outline of the methodology. In the subsequent sections, each step will be discussed in detail, in the context of our test case defects. 

The standard method for treating isolated point defects in semiconductors and insulators via DFT calculations is to construct a supercell with a large amount of host material to separate the defect from its periodic images \cite{Messmer1970,Louie1976,Freysoldt2014}. In this context, the goal of the quantum embedding approach for defects \cite{Bockstedte2018,Ma2020} is to treat the host semiconductor at the DFT level, while using a MB method (i.e., one that can handle the possibility of correlated, multi-determinant states) to treat the electronic structure of the defect. To do this, the Bloch states related to the defect are isolated from the bulk-like states, and transformed to a localized basis via Wannierization (see Sec.~\ref{sec:tijwan}). The defect states are
treated as a ``correlated subspace'' with the Hamiltonian
\begin{equation} \label{eq:MB}
\begin{split}
    H&=-\sum_{ij,\sigma}(t_{ij}c^\dagger_{i\sigma}c_{j\sigma}+ \text{H.c.})
    \\
    &+\frac{1}{2}\sum_{ijkl,\sigma\sigma^\prime}U_{ijkl} c^\dagger_{i\sigma}c^\dagger_{j\sigma^\prime}c_{l\sigma^\prime}c_{k\sigma}\\
    &-H_{\text{DC}}-\mu\sum_{i,\sigma} c^\dagger_{i\sigma}c_{i\sigma} ,
    \end{split}{}
\end{equation}
where $\sigma, \sigma'$ indicate spin and $i$,$j$,$k$,$l$ correspond to defect-related states; $t_{ij}$ are the hopping matrix elements between defect states in our Wannier basis (Sec.~\ref{sec:tijwan}); $U_{ijkl}$ are the Coulomb matrix elements in the correlated subspace, screened by the rest of the states in the supercell (Sec.~\ref{sec:cRPA}); $H_{\text{DC}}$ is a ``double-counting'' correction for the Coulomb interaction included in $t_{ij}$ (Sec.~\ref{sec:DC}); and $\mu$ is a chemical potential used to enforce the nominal occupation of the defect states. Only neutral excitations are considered here, i.e., we do not consider ionization of the defect. By utilizing  localized Wannier functions to describe the correlated subspace, we can restrict ourselves to the minimum of involved correlated states. The impurity-bulk ``connection'' is thereby established in two ways. First, the properties of the Wannier orbitals are controlled by the impurity geometry within the host material. Second, the Coulomb matrix elements are screened by the host environment. 

For the defects described in this work, the number of spin-orbitals in the correlated subspaces are quite modest, ranging from 4 for \cbcn{} to 10 for \feal{}. Thus, Eq.~(\ref{eq:MB}) can be exactly diagonalized (i.e., the full configuration interaction can be used \cite{Bockstedte2018,Ma2020}).
In the following sections we outline the methodology to determine the parameters in Eq.~(\ref{eq:MB}) and ultimately solve for the MB states.

Details of the computational parameters are provided in the SM \cite{SM} Sec.~S1.  
All DFT calculations are performed using the VASP code \cite{Kresse:1993bz,Kresse:1996kl,Kresse:1999dk}, and  Wannierization is performed via the interface to Wannier90 \cite{Mostofi2014}. The calculation of screened Coulomb matrix elements is calculated using the constrained random-phase approximation capabilities of VASP \cite{kaltak_merging_2015} (see Sec.~\ref{sec:cRPA}). The exact diagonalization is performed with tools in the {\sc triqs} \cite{triqs} library.

The strengths of this approach are two-fold. First, it leverages the extremely well-developed tools in the fields of DFT approaches for defects in semiconductors \cite{Freysoldt2014}, as well as solid-state embedding methods like DFT+dynamical mean-field theory (DMFT) \cite{Kotliar2006} and similar approaches \cite{Katsnelson1999}. Second, each step utilizes capabilities in widely available codes, which allows for simple reproduction and extension of our results and methodology.

\section{Methodological details and lessons from test cases \label{sec:lessons}}

In this section, we will expand upon various aspects of the embedding methodology outlined in Sec.~\ref{sec:method}, using the test-case defects introduced in Sec.~\ref{sec:defects}. In particular, we will focus on some basic aspects of the initial DFT calculation on which the embedding is based (Sec.~\ref{sec:initial}), including the choice of exchange-correlation functional (Sec.~\ref{sec:functional}) and the treatment of atomic relaxations in the ground and excited states (Sec.~\ref{sec:zpl}). We will discuss the downfolding procedure, i.e., the Wannierization (Sec.~\ref{sec:tijwan}) and calculation of the screened Coulomb interaction (Sec.~\ref{sec:cRPA}). Finally, we will discuss the Coulomb double counting (DC) problem resulting from joining DFT and exact diagonalization (Sec.~\ref{sec:DC}).

\subsection{Initial DFT electronic structure for defect geometries and the  noninteracting Hamiltonian} \label{sec:initial}

The role of the initial DFT calculation in the embedding procedure is three-fold: (i) to obtain an accurate atomic geometry for the defect; (ii) to provide the bulk band structure as needed to evaluate the screening of interaction matrix elements $U_{ijkl}$ in the active space (see Sec.~\ref{sec:cRPA}); and (iii) to provide the single-particle terms $t_{ij}$ of Hamiltonian in the active space given in Eq.~(\ref{eq:MB}). The challenge for embedding approaches for defects is to balance these aspects, i.e.,  obtaining an accurate electronic and atomic structure, as well as an appropriate noninteracting starting point for the embedding method.

For example, many defects are paramagnetic, i.e., they have nonzero spin. However, for the Hamiltonian in Eq.~(\ref{eq:MB}), the exchange interaction is accounted for in the interaction term (second line), and should not be included in the hopping matrix elements $t_{ij}$. Also, exchange-correlation (XC) functionals such as HSE \cite{HSE2003,HSE2006} are expected to result in more accurate band gaps \cite{Garza2016}, which would translate to more accurate bulk screening, and possibly more accurate defect geometries \cite{Freysoldt2014} but will also increase the Coulomb interaction effectively included in $t_{ij}$, which must be accounted for in the DC term in Eq.~(\ref{eq:MB}). The corresponding discussion on the impact of functional choices is given in Sec.~\ref{sec:functional}, where we will compare calculations with the PBE \cite{Perdew1996} semilocal generalized-gradient approximation (GGA) and HSE \cite{HSE2003,HSE2006} hybrid functionals.

In order to strike this balance between accurate structural parameters and a good noninteracting starting point, we will use the following procedure for the initial DFT calculations. First, atomic relaxations are performed using the standard procedure \cite{Freysoldt2014}, i.e., spin polarization is included and a finite $2\times2\times2$ $\Gamma$-centered $k$ mesh is used to preserve the symmetry in defects/hosts with hexagonal symmetry. This is intended to obtain an accurate defect structure. Afterwards, with the geometry fixed, an additional nonspinpolarized calculation is performed. From this calculation, we will obtain all necessary hopping and Coulomb interaction matrix elements for the MB Hamiltonian, as discussed in the next sections.

In certain cases, there may be different options for the initial spinless electronic configuration. Considering our case-study defects, \cbcn{} is completely unambiguous: since the single particle ground state is spin degenerate [two electrons in $b_2$, see Fig.~\ref{fig:cbcn_bs}(a)] the initial DFT calculation is identical whether or not spin is included. For \nv{}, the ground state contains a half-filled $e$ manifold, which would form a spin triplet in the spinfull case. In principle there are different options for constructing an initial spinless state, though the straightforward choice is one (spinless) electron in each $e$ orbital to preserve the symmetry. We find that the electronic structure of this state is very similar to the triplet state. For example, relaxing the geometry of the defect with or without spin results in identical structures (differences less than $3\times10^{-3}$ \AA{}).   

\feal{} in AlN is a case where including spin polarization in the DFT calculations significantly changes the electronic structure. It was shown in Ref.~\onlinecite{Wickramaratne2019} that there is a large spin splitting in the ground state of Fe$^{3+}$ in III-nitrides, whereas neglecting spin results in spin-degenerate states in the band gap of AlN [see Fig.~\ref{fig:feal_el_struct}(a)]. Since these states should be filled by five electrons, the $e$ states will be completely filled, and there will be one electron in the $t_2$ manifold. In the limit of a small thermal smearing, this results in $1/2$ of an electron in the two $e$ states that are slightly split from the $a_1$ state by the $C_{3v}$ crystal field. This is the electronic structure that we use as the first step in our calculations. One could also consider attempting to construct a spinless initial state that is closer to the spin-polarized structure, e.g., by constraining the occupation of the Fe $3d$ Kohn-Sham states such that all five are half filled. However, this constrained occupation is somewhat at odds with the spirit of Eq.~(\ref{eq:MB}), where the DFT calculations are intended to approximate a noninteracting  calculation.

\subsection{Structural relaxation in excited states for the zero-phonon line \label{sec:zpl}}

\begin{figure}
   \includegraphics[width=0.50\textwidth]{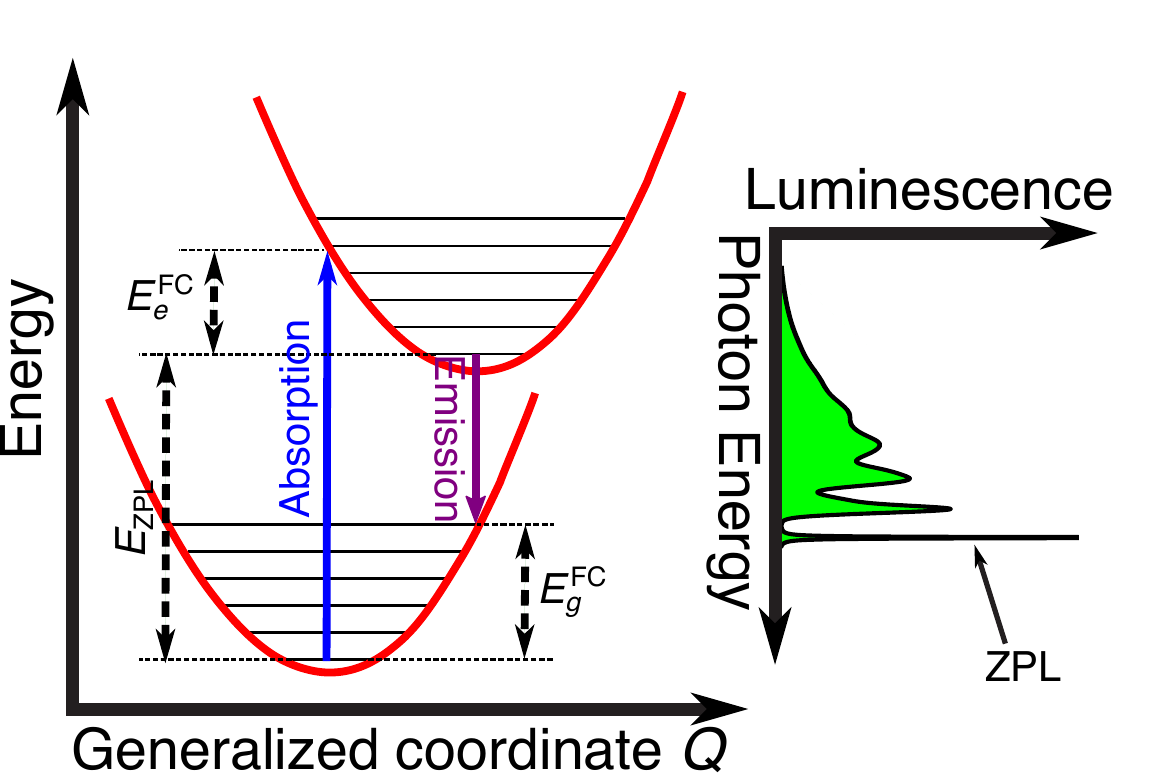}
\caption{\label{fig:CC} Schematic of the one-dimensional configuration-coordinate picture of optical absorption and emission transitions. The $y$ axis is energy, and the $x$ axis is a generalized coordinate that describes the coupling of the electronic energies of the defect (red curves) and the lattice. The zero-phonon line (ZPL) energy is labelled, as well as the Frank-Condon relaxation energies in the excited ($E^{\text{FC}}_e$) and ground ($E^{\text{FC}}_g$) electronic state. The inset shows a schematic of a luminescence spectra with the ZPL labelled.   }
\end{figure}

One of the key experimental observables from optical measurements is the ``zero-phonon line'' (ZPL) energy \cite{Stoneham,Davies}, which corresponds to a relatively sharp spectral line (for cases of weak to moderate electron-phonon coupling \cite{Alkauskas2016_Tut}) at the high-energy threshold for luminescence, or equivalently, the low-energy threshold for absorption (see Fig.~\ref{fig:CC}). The ZPL corresponds to the transition energy between the excited and ground state, each at their equilibrium atomic structure. Below, we will show ZPL results for \cbcn{} and \nv{} in order to compare with experiments and previous calculations.

At present, we do not have a way of performing atomic relaxations in the excited state, so we resort to the standard method \cite{Dreyer2018} of performing a constrained DFT (cDFT) calculation to approximate the electronic structure of the excited state, and then relaxing the atoms under the constraint. Once we have an approximation for the structure of the excited state, we then find the difference between the vertical transition (i.e., the difference in MB energies) calculated for the ground and excited state structures $\Delta E_{\text{MB}}$ (i.e., the difference in the vertical absorption and emission energies in Fig.~\ref{fig:CC}). This gives us the sum of the Frank-Condon (FC) relaxation energy in the excited and ground states, $E^{\text{FC}}_g+E^{\text{FC}}_e$ (see Fig.~\ref{fig:CC}). We assume that these relaxation energies are equal, and thus $E^{\text{FC}}_g=E^{\text{FC}}_e=\Delta E_{\text{MB}}/2$, which is quite accurate for \cbcn{} \cite{Mackoit2019} and \nv{} \cite{Alkauskas2014}. 

For \cbcn{}, we are interested in the $\vert \text{GS} \rangle\rightarrow\vert \text{D} \rangle$ transition, since it is the one attributed to the 4.1 eV ZPL in experiment \cite{Mackoit2019}. Approximating the excited state $\vert \text{D} \rangle$ is quite straightforward, we just populate the $b_2^*$ antibonding orbital with one electron taken from the $b_2$ bonding orbital.
The key observable for the \nv{} center is its 1.945 eV ZPL corresponding to transitions between $^3A_2$ and $^3E$ \cite{Doherty2013}, which allows for a quantitative test of the methodology. The excited state triplet $^3E$ is approximated by promoting an electron from $a_1(2)$ to the $e$ manifold [see Fig.~\ref{fig:NV_elec_struct}(a)]. In order to avoid the Jahn-Teller distortion and remain in the $C_{3v}$ symmetric structure, the electron density is spread between the $e$ states \cite{Alkauskas2014}.

We note that this approach, i.e., relying on cDFT for the excited-state structure, and the fact that $E^{\text{FC}}_g=E^{\text{FC}}_e$, may not be generally applicable, and thus motivates the implementation of forces in the embedding scheme, which will be the topic of future work. 

\subsection{Downfolding via Wannierization \label{sec:tijwan}}

The next step of the calculation is to construct the active space and to ``downfold'' the KS space to this subspace.
As discussed in Sec.~\ref{sec:initial}, with the ground-state geometry fixed, we perform a \emph{spinless} DFT calculation. From this we construct the localized basis for the correlated subspace $\phi_i(r)$ via Wannier constructions utilizing the Wannier90 \cite{Mostofi2014} package.
In the cases where the correlated defect states are in the gap, such as \cbcn{} and \feal{} we surround them with a ``frozen window'' so that the single-particle Wannier Hamiltonian defined by the hopping matrix elements
\begin{align}
    t_{ij} = -\langle \phi_i \vert H_{\text{DFT}} \vert \phi_j \rangle
\end{align}
reproduces the DFT eigenvalues exactly; for states that are resonant with the bulk bands, such as the lower $a_1(1)$ state in \nv{}, we rely on initial projections of defect orbitals to disentangle the defect states. 

Due to the gauge freedom when constructing Wannier functions, it is crucial to ensure that the specific procedure to generate them does not influence the final results. To test this, we perform calculations of the MB states of \cbcn{} with and without localizing the Wannier functions from their initial projections (on C $p_z$ orbitals), but we find that the final MB energies are identical, confirming that the gauge choice does not modify the observables.

In the case of the \nv{} center, the Wannierization procedure is slightly more complicated since, as mentioned above, the $a_1(1)$ state must be disentangled from the valence band manifold. To do this, we chose $sp^3$ initial projections on the atoms surrounding the vacancy [see Fig.~\ref{fig:NV_elec_struct}(b)], but, unlike for \cbcn{}, we maximally localize the Wannier functions. The localization procedure is constrained to exactly reproduce the Kohn-Sham eigenvalues for the $a_1(2)$ and $e$ states in the gap, with a disentanglement window large enough to include the $a_1(1)$ state in the diamond VB. In any case, the relevant MB excited states do not involve significant depopulation of $a_1(1)$, so the specifics of its treatment is not so crucial. We find that this procedure provides us with an accurate basis for subsequent MB calculations.

The situation for \feal{} is similar to \cbcn{} in that all of the relevant states are in the band gap [Fig.~\ref{fig:feal_el_struct}(a)]. Thus, if we begin with $d$ projections on the Fe, there is very little change whether we localize or not. As with \cbcn{}, we will not localize in order to preserve the symmetry of the basis. 

\subsection{Obtaining the screened interaction parameters \label{sec:cRPA}}

\subsubsection{Constrained random-phase approximation}

The next step is to obtain screened interaction parameters $U_{ijkl}$ in the subspace of defect orbitals. We construct these from the localized Wannier basis via 
\begin{align}
    U_{ijkl}&= \braket{\phi_i  \phi_j \vert \widehat{U} \vert \phi_k  \phi_l } \\
            &= \int \int d^3r \, d^3r' \, \phi_i^*(r) \phi_k(r) \, U(r,r') \, \phi^*_j(r') \phi_l(r') \notag
\end{align}
using the partially screened Coulomb interaction in the static limit
\begin{align}\
\label{eq:screen_U}
    \widehat{U} = \left[ 1- \hat{v} \, \widehat{\Pi}_\text{cRPA}(\omega=0) \right]^{-1} \hat{v}.
\end{align}
Here $\hat{v}$ is the bare Coulomb interaction and $\widehat{\Pi}_\text{cRPA}$ is the partial polarization as defined within the constrained random phase approximation (cRPA) as~\cite{Aryasetiawan2004}
\begin{align}
    \widehat{\Pi}_\text{cRPA} = \widehat{\Pi}_\text{full} - \widehat{\Pi}_\text{defect},
\end{align}
where the ``full'' polarization takes all RPA screening processes from the KS states into account, and the ``defect'' polarization accounts only for screening processes within the defect-state manifold. In this way $U_{ijkl}$ is screened by the bulk host material; the screening \emph{within} the defect-state manifold is subsequently included exactly via the solution of the Hamiltonian defined in Eq.~(\ref{eq:MB}). We perform these cRPA calculations using a recent implementation by Kaltak \cite{kaltak_merging_2015} within VASP. This method requires a mapping between the Wannier and Bloch-band bases to define $\widehat{\Pi}_\text{defect}$; this mapping is exact if no disentanglement is necessary, and we find that, in any case, the results are insensitive to the specific method used (i.e., ``weighted'' \cite{sasiolgu_effective_2011} versus ``projected'' \cite{kaltak_merging_2015}). 

Finally, we stress that RPA and cRPA calculations based on KS-DFT input for gapped systems benefit from an error cancellation initially introduced due to missing higher order diagrams in RPA \cite{vanLoon2021}. 

\subsubsection{Convergence of the screened interaction \label{sec:conv_u}}

\begin{figure}
   \includegraphics[width=\columnwidth]{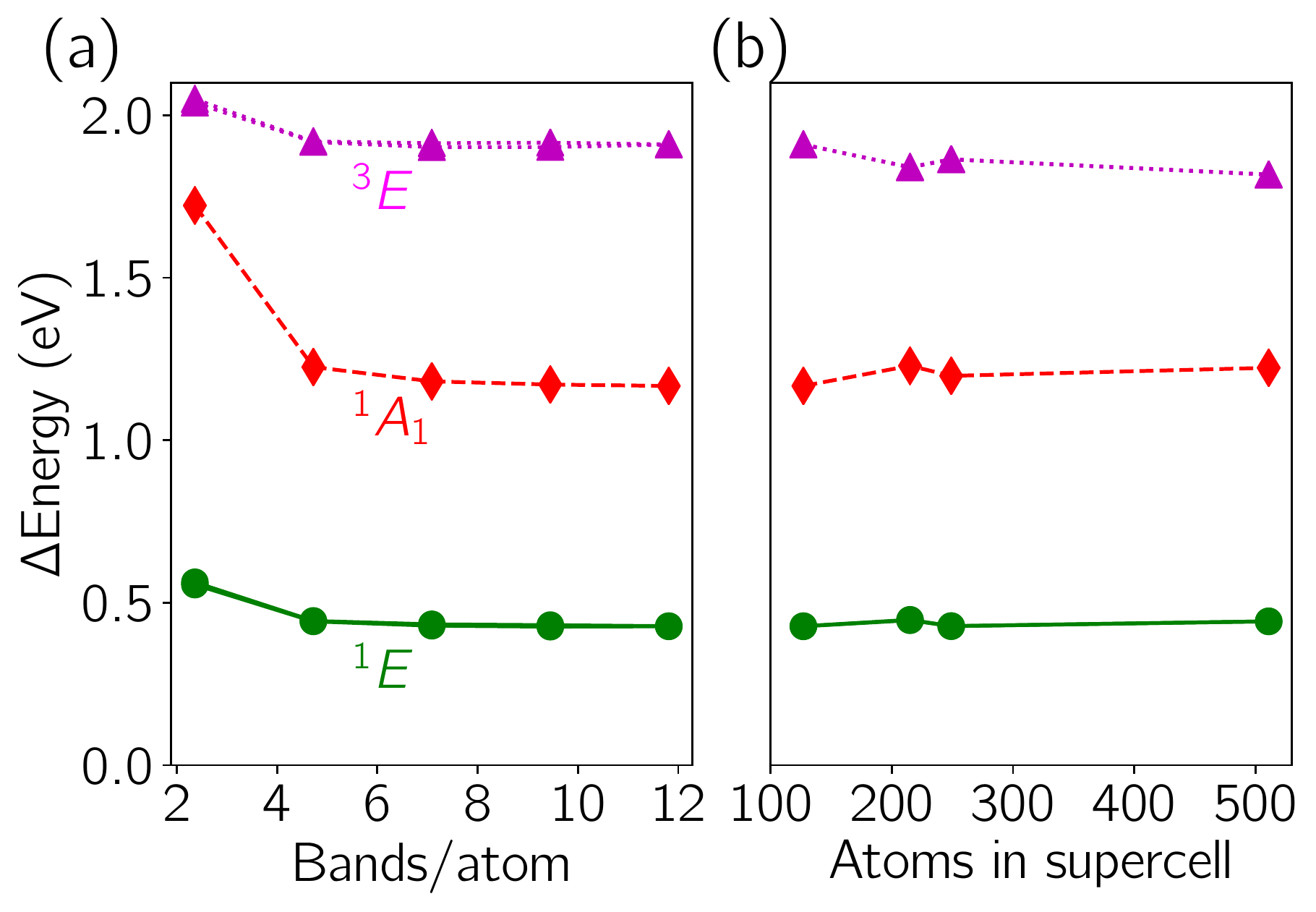}
\caption{\label{fig:NVbench} Convergence of the \nv{} energies of many-body states with (a) number of bands in the cRPA calculation (for a 127 atom cell), and (b) supercell size (Bands/atom $>7$ for all cells). The PBE functional is used, and no double counting correction is applied.}
\end{figure}

The convergence of the screening is key to obtaining accurate MB energies, as we will demonstrate for our test-case defects.
In the bulk, the screening should be converged with respect to the number of virtual orbitals (in conventional sum-over-states implementations), and $k$ points used to sample the Brillouin Zone. For the defective system, increasing the supercell size with a single $k$ point achieves the latter convergence via band folding, while maintaining the zero-dimensional (0D) nature of the calculation (i.e, avoiding contributions from the spurious dispersion of the defect states caused by interactions between periodic images). In Fig.~\ref{fig:NVbench}, we show that the convergence of the excited-state energies for \nv{} (with respect to the $^3A_2$ ground state). We see that the main source of convergence is the number of empty bands used in the cRPA calculation; the influence of supercell size is relatively less important. This is a significant result, since it is often found (e.g., Ref.~\onlinecite{Bockstedte2018}) that large supercells ($>500$ atoms) are necessary to obtain converged energies for \nv{}; we believe that our localized basis set is the reason for the significantly improved convergence with supercell size.

\begin{figure}
   \includegraphics[width=\columnwidth]{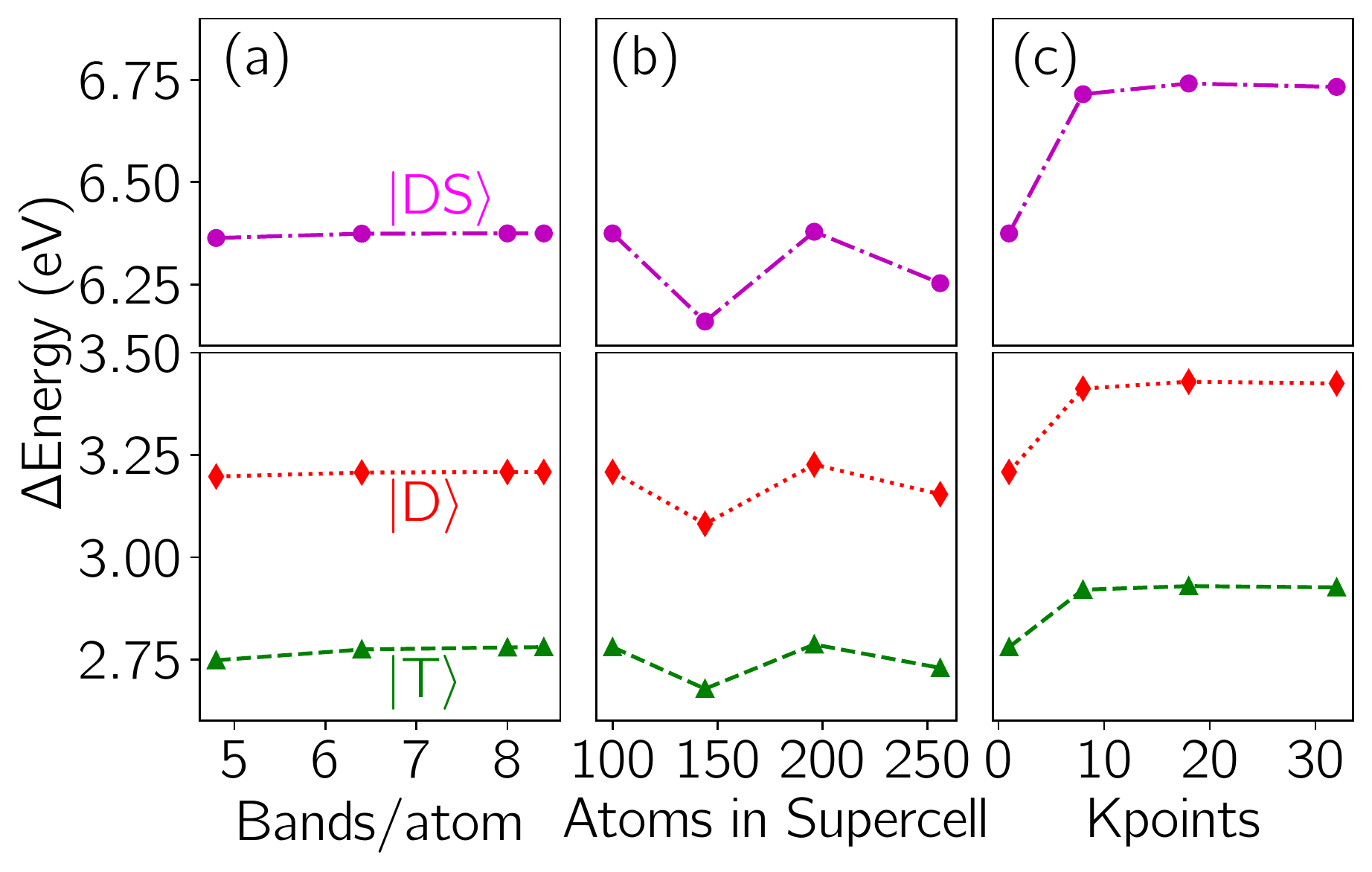}
\caption{\label{fig:cbcn_conv} Convergence of the \cbcn{} energies (referenced to the ground-state singlet) of the triplet $\vert \text{T}\rangle$, first excited singlet $\vert \text{D}\rangle$, and second excited state singlet $\vert \text{DS}\rangle$ with respect to (a) bands/atom (in the 100 atm cell), (b) size of supercell and (c) number of in-plane $k$ points. The PBE functional is used and no double-counting correction has been applied.}
\end{figure}

\begin{figure}
   \includegraphics[width=0.8\columnwidth]{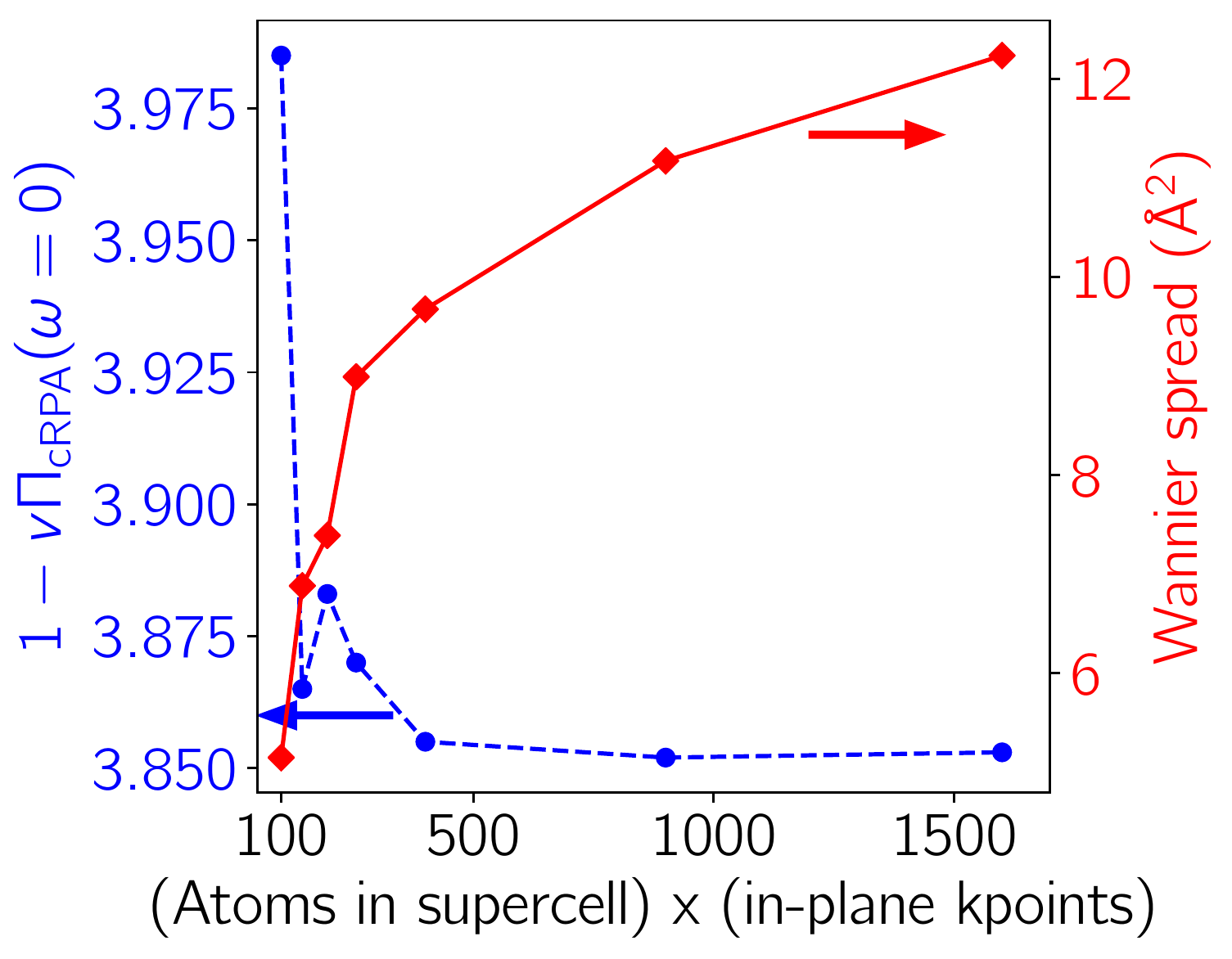}
\caption{\label{fig:cbcn_screen}  Convergence of the polarizabilty [see Eq.~(\ref{eq:screen_U})] and spread of the Wannier functions for \cbcn{} with respect to effective atom number, i.e., the number of atoms in the super cell multiplied by the number of $k$ points in plane. Calculations are for projected Wannier functions with C $p_z$ character. The PBE functional is used.} 
\end{figure}

The convergence in the case of \cbcn{} is more complicated. As with \nv{}, the MB excitation energies are well-converged for $>5$ bands per atom in the supercell [Fig.~\ref{fig:cbcn_conv}(a)]. However, we see from Fig.~\ref{fig:cbcn_conv}(b) that the convergence with supercell size is difficult to achieve; for the accessible supercell sizes (before we are limited by the computational demand of the cRPA calculation), the MB energies have oscillatory behavior. The reason for this is the slow convergence of the screened Coulomb interaction with cell size, as shown via the polarizability [see Eq.~(\ref{eq:screen_U})] in Fig.~\ref{fig:cbcn_screen}. We find that an effective method for accelerating this convergence, is to increase the in-plane $k$-point mesh, as shown in Figs.~\ref{fig:cbcn_screen} and \ref{fig:cbcn_conv}(c). We can think of the number of atoms multiplied by the number of in-plane $k$ points as an ``effective'' supercell size from the point of view of the bulk screening. From Fig.~\ref{fig:cbcn_conv}(b), the largest supercells we could treat contained 250 atoms, while the screening in this layered compound clearly requires at least double that to converge, likely due to the non-local character of the bulk background dielectric function of layered semiconductors \cite{Huser2013,Andersen2015,Rosner2015}. In addition, we see from Fig.~\ref{fig:cbcn_screen} that the spread of the Wannier functions also converges slowly with cell size. Fig.~\ref{fig:cbcn_conv}(c) shows that indeed, increasing the $k$ mesh allows us to obtain converged energies of the MB states.

Using multiple $k$ points means that our Wannier Hamiltonian is no longer strictly 0D, i.e., interdefect hopping is possible, and the spurious dispersion of the defect states is sampled. However, we find that even for the $5\times5\times1$ cell, the largest intersite hopping element is 0.09 eV \footnote{This hopping is out of plane, and can be reduced to 3 meV by doubling the cell in the $c$ direction. However, we find that using a larger cell in $\hat{c}$ does not significantly change the MB energies ($<20$ meV).}. This gives us confidence that we are not affected by the spurious dispersion caused by interactions between defects and their periodic images.

\begin{figure}
   \includegraphics[width=0.50\textwidth]{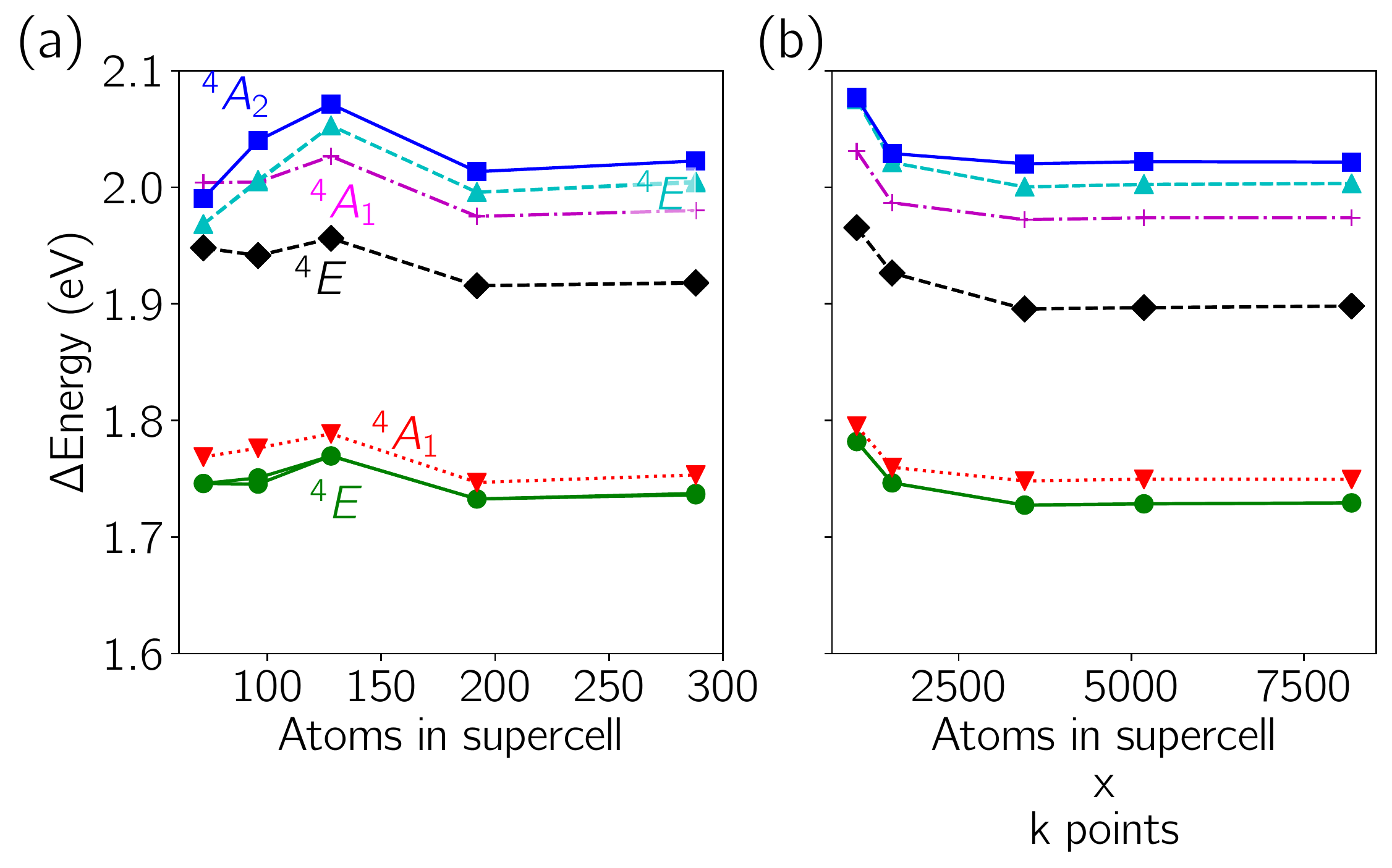}
\caption{\label{fig:AlNbench} Convergence of energies of many-body states of \feal{} in wurtzite AlN referenced to the $^6A_1$ with (a) number of atoms in the supercell (Bands/atom $>11$ for all cells); (b) ``effective'' supercell size, i.e. atoms multiplied by $k$ points; points correspond to a 128 atom and 192 atom cells with a $2\times2\times2$, $3\times3\times3$, and $4\times4\times4$ (only for the 128 atom cell). The PBE functional is used and no double-counting correction has been applied.  }
\end{figure}

We find in the case of \cbcn{} that the out-of-plane $k$ mesh and supercell size has a relatively small (see SM \cite{SM} Sec.~S1 A) effect on the screening of the MB states, which is due to the quasi 2D nature of the host material BN. However, we can use the same procedure for a 3D material, but increasing the $k$ mesh in all three dimensions. We demonstrate this for our remaining test case, \feal{}, in Fig.~\ref{fig:AlNbench}. In panel (a) we show the convergence of the energies of excited MB states with respect to supercell size. Though the convergence can be achieved, a cell of around 200 atoms is required for quantitative accuracy. In Fig.~\ref{fig:AlNbench}(b) we show this convergence versus ``effective'' supercell size, i.e., the number of atoms multiplied by the number of $k$ points (in all directions this time since our system is 3D). 
The points at large effective size correspond to 128-atom and 192-atom cells with a $2\times2\times2$, $3\times3\times3$, and $4\times4\times4$ (only for the 128 atom cell) $k$ meshes. As we can see the MB energies converge smoothly, agreeing with the results from larger actual supercells. The largest hopping between the defect and its periodic images was less than 0.006 eV, indicating that we have preserved the 0D nature of our noninteracting Wannier Hamiltonian.

\subsection{Choice of DFT functional for initial calculation \label{sec:functional}} 

We now turn to another crucial element relating to the embedding procedure, the choice of XC functional for the DFT calculation. To explore this, we compare calculations with the HSE hybrid functional \cite{HSE2003,HSE2006}, which has become the standard for quantitative calculations of defect properties \cite{Freysoldt2014} to the results with PBE \cite{Perdew1996}, which is one of the most popular functionals for solid-state applications.

A key question regarding the use of hybrids like HSE is the choice of mixing parameter, $\alpha$, since gaps between single-particle levels (e.g., the band gap of the host material) scales with $\alpha$. In many cases, $\alpha$ is chosen to roughly reproduce the experimental band gap of the material, though there are also \textit{ab-initio} approaches to determining the mixing \cite{Skone2014}. In this paper, we will rely on the values shown to produce accurate results in previous computational studies. We will show in Sec.~\ref{sec:DC} that an appropriate double-counting correction for hybrids should include $\alpha$, and thus should remove much of the dependence of the MB energies on the starting XC functional.

For the case of \cbcn{}, we tune the mixing parameter of the HSE functional to $\alpha=0.4$, as was done in Ref.~\onlinecite{Mackoit2019}. The main quantitative effect of HSE is on the single-particle states. Firstly, the eigenvalue difference between the C-derived states increases by a factor of 1.5 (from $3.45$ eV with PBE to $5.25$ eV with HSE), as does the band gap ($4.54$ eV to $6.85$ eV). 
Though the Wannier-function spread is slightly reduced with HSE, which results in a slightly larger unscreened Coulomb interaction, the main effect on the $U_{ijkl}$ elements is due to the reduced environmental screening (due to the larger gaps between single-particle states, both bulk-bulk and bulk-defect). This results in a significant increase in the screened Coulomb matrix elements, e.g., of more than 500 meV for the density-density terms. If we perform an average over the orbitals in the Wannier basis (see Sec.~\ref{sec:models}), we obtain intraorbital $U=2.73$ eV, interorbital $U^\prime=1.90$ eV, and Hund's coupling $J=0.09$ eV. Compared to the values we calculated for PBE ($U=1.94$ eV, $U^\prime=1.41$ eV, and $J=0.08$), the most significant change is an increase in the intraorbital $U$.

For the MB states, the effect of HSE is manifested as an increase in energy between $\vert\text{GS}\rangle$ and the excited state singlet $\vert\text{D}\rangle$ of $1.62$ eV. The splitting between $\vert\text{T}\rangle$ and $\vert\text{D}\rangle$ only increases by $0.12$ eV, as it depends on the exchange interaction and not the splitting of the single-particle levels. The  $\vert\text{GS}\rangle-\vert\text{DS}\rangle$ splitting is increased by 3.26 eV, or approximately twice the increase in $\vert\text{GS}\rangle-\vert\text{D}\rangle$, due to the fact that $\vert\text{DS}\rangle$ involves two electrons in the antibonding orbital. See the ``No DC'' points on Fig.~\ref{fig:cbcnDC_HSE}(a) and (b) for a comparison of the energies calculated with HSE and PBE. 

Compared to the $\Delta$SCF HSE calculations of Ref.~\onlinecite{Mackoit2019}, the singlet-singlet splitting ($\vert\text{GS}\rangle-\vert\text{D}\rangle$) energies that we obtain are about 500 meV larger; e.g., the ZPL that we find is 4.86 eV versus 4.31 eV in Ref.~\onlinecite{Mackoit2019} [c.f., ``No DC'' in Fig.~\ref{fig:cbcnDC_ZPL}(a) and (b)]. The reason for this overestimation is  least partially be due to the fact that, as discussed in Sec.~\ref{sec:method}, the original DFT calculation includes some approximate Coulomb interaction, the effect of which should be removed with the DC correction (discussed in the next section). Hybrid functionals often have a similar effect on the electronic structure in terms of improving the description of localized states as, e.g., DFT+$U$, and thus we expect that they contain more of the Coulomb interaction than local and semilocal functionals. 

The HSE functional has been used in the past to obtain optical properties of \nv{} in excellent agreement with experiment \cite{Weber2010,Alkauskas2014,Gali2009_PRL}. The main effect of the HSE ($\alpha=0.25$) functional for \nv{} (similar to \cbcn{}),  is to increase the splitting between the $a_1(2)$ and $e$ levels (2.14 eV for HSE versus 1.50 eV for PBE at the gamma point for the spinless initial calculation). This results in a significantly increased splitting between the ground state triplet $^3A_2$ and the excited state triplet $^3E$ (2.73 eV vertical excitation for HSE versus 1.84 for PBE, see ``No DC'' points in Fig.~\ref{fig:NV_DC}), as well as, to a lesser extent, $^1A_1$ (1.39 eV vertical excitation for HSE versus 1.22 eV for PBE), both of which involve exciting an electron from the $a_1$ to the $e$ state. The energy of the $^1E$ state changes less between PBE and HSE (0.54 eV vertical excitation for HSE versus 0.45 eV for PBE), as it is not directly influenced by the splitting of single-particle states. Comparing the Coulomb tensors, the main difference between PBE and HSE is that HSE has a slightly larger intraorbital screened interaction (averaged parameters for HSE: $U=2.83$ eV, $U^\prime=0.98$, $J=0.02$ versus $U=2.43$ eV, $U^\prime=0.81$ eV, and $J=0.03$ eV for PBE, see Sec.~\ref{sec:models}). This is a result of the decreased environmental screening due to the larger band gap in HSE. 

Thus, from the examples of \cbcn{} and \nv{}, we would conclude that HSE provides a better description of the initial single-particle electronic structure (e.g., bulk band gaps and splitting of single particle levels), and thus is a good starting point for our embedding methodology. However, the case of \feal{} is significantly more complicated. The reason for this, as pointed out above, is that HSE includes additional aspects of the interaction (i.e., exact exchange) in the part of the Hamiltonian that is intended to be noninteracting.

For HSE, the $a_1$ state that is split from $t_2$ by the $C_{3v}$ crystal field is significantly lower in energy than the $e$ state (by 1.5 eV), and there is a sizeable splitting in the $e$ state single particle levels of 200 meV (in spite of the fact that the calculation has $C_{3v}$ symmetry). If we constrain the electronic structure to look more like that calculated with PBE, i.e., forcing the $e$ states to be half filled, we recover the degeneracy of the $e$ state by construction. The $a_1$ state is now \emph{higher} in energy by 900 meV, compared to 20 meV for PBE. This difference cannot be attributed to structural difference between HSE and PBE, as performing a PBE calculation with the HSE structure give an electronic structure that is close to PBE (i.e., the crystal field splitting of the $t_2$ state is 30 meV). 

The large splitting in the ground state results in MB states that significantly differ from the PBE results, experimental observations \cite{Malguth2006,Malguth2008,Neuschl2015}, and spinful DFT calculations \cite{Wickramaratne2019}. Specifically, the ground state is low spin ($S=1/2$), as predicted \cite{Sugano1970} for large CFS, as opposed to the high spin $S=5/2$ as expected \cite{Malguth2006,Malguth2008,Neuschl2015,Wickramaratne2019} (see ``No DC'' points on Fig.~\ref{fig:feal_DC}). Thus, though HSE clearly provides a better description of the bulk electronic structure and screening in AlN (and the other host materials discussed in this work), it appears problematic in this case for a noniteracting starting point of the correlated subspace. We will extend on this discussion after introducing the DC correction in the next section.

\subsection{Double-counting correction} \label{sec:DC}

In the previous sections, we discussed how we obtain the single particle and the screened interaction matrix elements in the correlated subspace as needed for the Hamiltonian in Eq.~(\ref{eq:MB}). In principle, however, the separation of the noninteracting part from the Coulomb interactions has fundamental incompatibilities with DFT calculations. This is because the DFT calculation already includes Coulomb interactions within the correlated subspace in an approximate way, which does not have a rigorous definition within MB perturbation theory \cite{Kotliar2006, Haule2015}. This issue is usually dealt with by applying a ``double counting'' (DC) correction to the hopping matrix elements. 

Within the DFT-based embedding community [i.e., DFT+$U$ and DFT+dynamical mean-field theory (DMFT)], the most common approaches apply a DC correction potential that involves orbitally-averaged interaction parameters $U$ and $J$, and moreover assumes no orbital polarization, i.e., that the orbital levels are degenerate \cite{Liechtenstein1995,Haule2015}. Such a DC correction will shift the correlated subspace with respect to the uncorrelated one, and possibly alter the total occupation, but will not change the splitting between orbitals in the correlated subspace, which is our focus.
Also, in our work, the occupation is enforced in all cases in the MB calculation via the chemical potential $\mu$ in Eq.~(\ref{eq:MB}). Thus a fully orbitally-averaged DC correction will have no effect on the results.

It has been found before \cite{Nekrasov2012,Kristanovski2018,Bockstedte2018,Ma2020} and is confirmed in our results below that an \textit{orbitally selective} version of the DC correction is required to obtain agreement with experiment.
However, there is vanishingly little work on benchmarking such an approach in general, and systematic investigations of the DC for defect embedding methodologies are not available, yet.

\subsubsection{Form of the orbitally-resolved double-counting correction}

A systematic methodology for obtaining an orbitally-resolved DC correction for extended systems was given in Ref.~\onlinecite{Haule2015}, and we will center our discussion around that approach. The general idea is to determine the DFT treatment of the Coulomb interaction in the subspace by making equivalent DFT approximations within that space. Thus, the charge density distribution is constructed from our Wannierized defect states, as opposed to the KS bands of the entire system. Also, the Coulomb interactions that enter the Hartree term, and are used to construct the XC, should be the screened interaction determined by cRPA (see Sec.~\ref{sec:cRPA}). 

The Hartree term is easy to calculate in our basis, and can be written as \cite{Bockstedte2018,Ma2020,Szabo1996}
\begin{equation}
\label{eq:DC_hartree}
    H^{\text{Har}}_{\text{DC}}=\sum_{ij,\sigma}c^\dagger_{i\sigma}c_{j\sigma} \sum_{kl}P_{kl} U_{iljk},
\end{equation}
where $P_{kl}$ is the component of the single-particle density matrix for (Wannier) orbitals $k$ and $l$ and $U_{iljk}$ is the density-density screened Coulomb interaction in our subspace. The XC term is more complicated to obtain \cite{Haule2015}; for, e.g., DFT under the local density approximation (LDA), it would require calculating the density-dependent XC energy of the uniform electron gas, using the screened interaction given by Eq.~(\ref{eq:screen_U}). An important point to note is that such a potential would be calculated from the \emph{total orbital averaged density} in the active space, as opposed to the Hartree term in Eq.~(\ref{eq:DC_hartree}), which is obtained from the \emph{orbitally specific} density-density interaction. Thus the XC part is expected to have a weaker contribution to relative shifts of orbitals within the active space than the Hartree term, although it will still have some influence since the different orbitals have different spatial distributions of charge densities. In this study, we will neglect this term, as was done in Ref.~\onlinecite{Bockstedte2018}. It is important to note that if we were interested in the alignment between the defect states of our correlated subspace and the bulk states, the XC contribution must be included for a consistent DC scheme. 

In the case where we are using a hybrid functional, we will have an additional orbital dependence arising from the Fock exact-exchange operator. For a full Hartree-Fock calculation, the Coulomb interaction, and thus the DC correction that should be subtracted in the MB Hamiltonian is \cite{Bockstedte2018,Ma2020,Szabo1996}:
\begin{equation}
\label{eq:DC}
    H^{\text{HF}}_{\text{DC}}=\sum_{ij,\sigma}c^\dagger_{i\sigma}c_{j\sigma} \sum_{kl}P_{kl}\left(U_{iljk}-\frac{1}{2} U_{ilkj}\right).
\end{equation}
This was used as the DC correction in Ref.~\onlinecite{Ma2008,Ma2020,Ma2020_2}. 
From Eqs.~(\ref{eq:DC_hartree}) and (\ref{eq:DC}), we see that a logical form of the DC correction for hybrid functionals is 
\cite{Bockstedte2018}:
\begin{equation}
\label{eq:DC_hyb}
    H^{\text{hyb}}_{\text{DC}}=\sum_{ij,\sigma}c^\dagger_{i\sigma}c_{j\sigma} \sum_{kl}P_{kl}\left(U_{iljk}-\alpha U_{ilkj}\right),
\end{equation}
where $\alpha$ is the mixing parameter of exact exchange in the hybrid functional. We will explore these forms of the DC in the next section.

\subsubsection{Effect of double-counting correction on many-body energies}

We will now discuss the orbitally-resolved DC corrections introduced in the previous section in the context of our test-case defects, starting with \cbcn{}. Because of the simplicity of the defect electronic structure, the DC correction is particularly simple to interpret in this case. In addition, the correspondence with the dimer model (see SM \cite{SM} Sec.~S2 A 1) allows the calculation of a ``Dimer'' DC that takes into account specifically the Coulomb interactions included in Kohn-Sham DFT \cite{vanLoon2021,Carrascal2015} for a system with a single filled valence orbital and empty conduction orbital [see Fig.~\ref{fig:cbcn_bs}(a)]. Note that, in principle,  this DC correction is only exact for the exact XC functional \cite{vanLoon2021}. A discussion of this form of the DC is given in the SM \cite{SM} Sec.~S2 A 2.

\begin{figure}
   \includegraphics[width=\columnwidth]{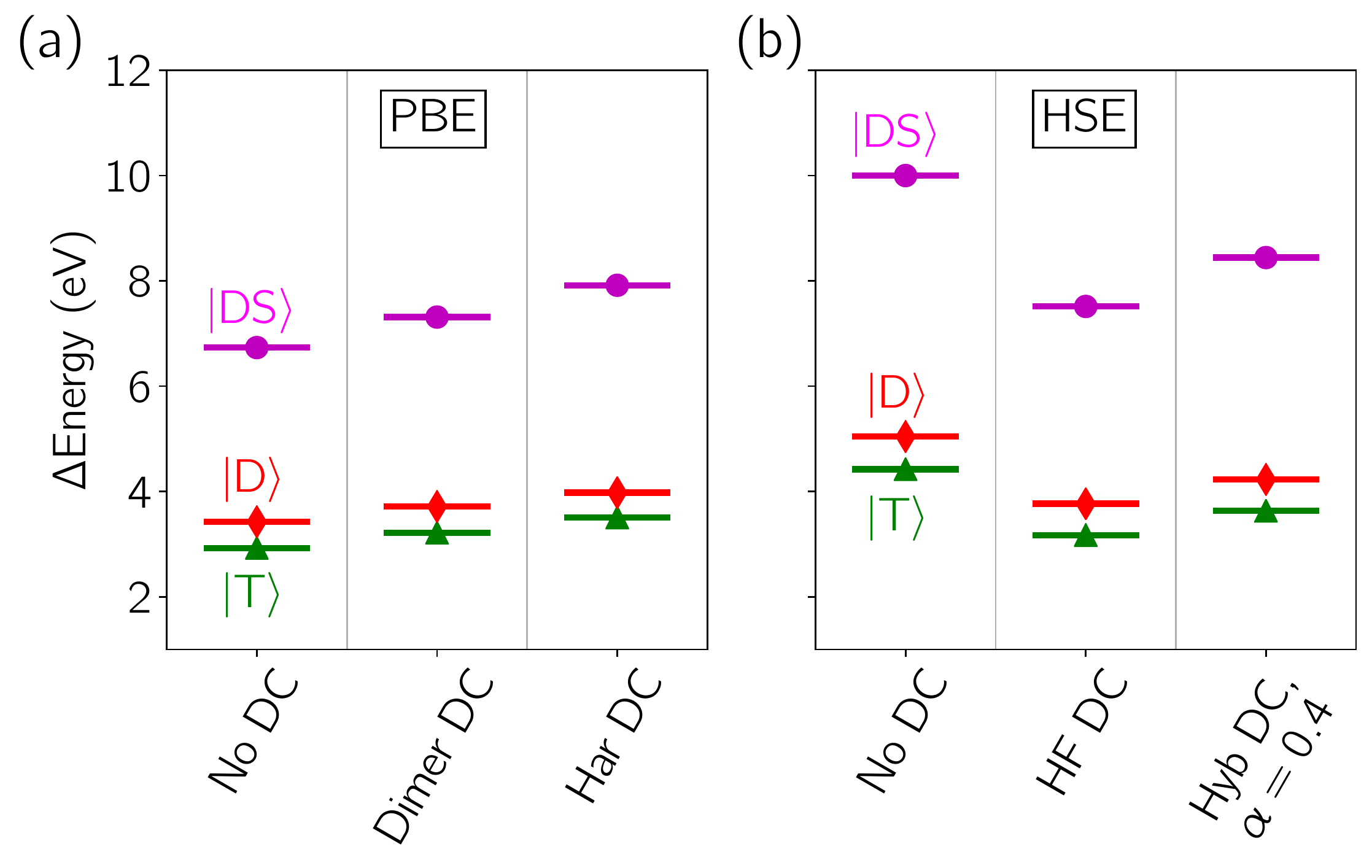}
\caption{\label{fig:cbcnDC_HSE} Excited state energies of \cbcn{} in BN with respect the ground-state singlet state calculated with either (a) PBE or (b) HSE, with and without double counting (DC) corrections (described in Sec.~\ref{sec:DC}).}
\end{figure}

Figure~\ref{fig:cbcnDC_HSE} shows the effect of the DC correction on the MB excitation energies for \cbcn{}. The levels on the left of each panel are calculated with no DC correction. For PBE [Fig.~\ref{fig:cbcnDC_HSE}(a)], we use the dimer DC (SM \cite{SM} Sec.~S2 A 2), and the approach of removing just the Hartree term [Eq.~\ref{eq:DC_hartree}]. In both cases, the main result of the DC correction is to increase the splitting between the $b_2$ and $b_2^*$ single-particle states; since all excited state involve electron(s) being promoted from $b_2$ to $b_2^*$, the energies of all excited states are shifted up in energy with respect to the ground state. The dimer DC involves an explicit shift of the levels (i.e., it is diagonal in the band basis by construction) of $\Delta\epsilon^{\text{DFT}}=U_{b_2b_2b_2b_2}-U_{b^*_2b_2b^*_2b_2}$ (SM \cite{SM} Sec.~S2 A 2). Since the intraorbital interaction is slightly larger than the interorbital one, the DC correction slightly increases the splitting between the single particle states. For $H_{\text{DC}}^{\text{Har}}$ (``Har DC'' on Fig.~\ref{fig:cbcnDC_HSE}), the only nonzero element of the density matrix is $P_{b_2b_2}=2$, so the DC correction shifts down the $b_2$ level, increasing the gap to $b_2^*$ (the shift is slightly larger than for the dimer DC due to small nonzero terms in the interaction in addition to the strict density-density $U_{b_2b_2b_2b_2}$ and $U_{b^*_2b_2b^*_2b_2}$).

In Fig.~\ref{fig:cbcnDC_HSE}(b) we compare the effect of the DC correction [full Hartree-Fock DC Eq.~(\ref{eq:DC}) and the hybrid version Eq.~(\ref{eq:DC_hyb})] on the MB energies using the HSE functional. We see that the DC has the opposite effect, the energies of the excited states are \emph{reduced} compared to the ground state. This is a direct result of the inclusion of exact exchange in the DC, which reduces the splitting between the single-particle defect levels. In the band basis, this change in splitting is given by $\Delta \epsilon_{\text{DC}}=U_{b_2b_2b_2b_2}-2U_{b^*_2b_2b^*_2b_2}+U_{b^*_2b_2b_2b^*_2}$. We find that, in our calculations, $U_{b_2b_2b_2b_2}\simeq U_{b^*_2b_2b^*_2b_2} \gg U_{b^*_2b_2b_2b^*_2}$ (e.g., for HSE: $U_{b_2b_2b_2b_2}=2.096$ eV, $U_{b^*_2b_2b^*_2b_2}=2.099$ eV, $U_{b^*_2b_2b_2b^*_2}=0.311$ eV). Therefore $\Delta \epsilon_{\text{DC}}<0$. In the $H_{\text{DC}}^{\text{Hyb}}$ case, this decrease is reduced due to the mixing parameter.

\begin{figure}
   \includegraphics[width=\columnwidth]{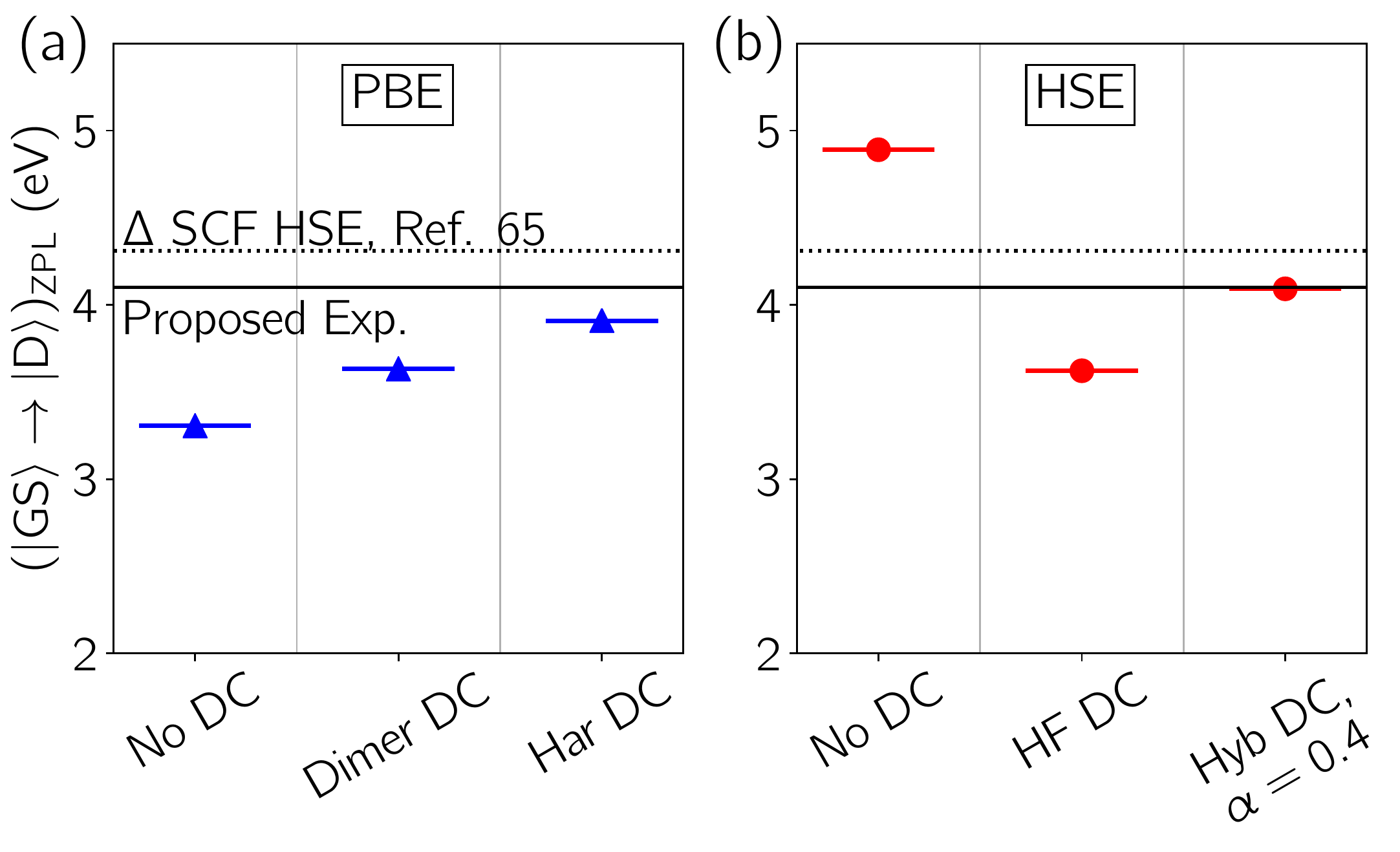}
\caption{\label{fig:cbcnDC_ZPL} (a) Singlet-singlet zero-phonon line energy of \cbcn{} in BN calculated using either (a) PBE or (b) HSE, with and without double counting (DC) corrections (described in Sec.~\ref{sec:DC}). Proposed experimental attribution and $\Delta$SCF HSE from Ref.~\onlinecite{Mackoit2019}. 
}
\end{figure}

In Fig.~\ref{fig:cbcnDC_ZPL}, we show the effect of the DC correction on the lowest-energy singlet-singlet ZPL (see Sec.~\ref{sec:zpl}) for \cbcn{}. We see that, in both cases, the DC improves the agreement with the $\Delta$SCF results of Ref.~\onlinecite{Mackoit2019}, and thus the proposed experimental attribution. 

We also note that using an appropriate DC correction, i.e., ``Dimer'' or ``Har'' for PBE and ``Hyb'' for HSE, significantly reduces the dependence of the final results, in terms of MB energies and ZPL, on the XC functional used for the DFT starting point. As this was the intended role of the DC correction, these results are quite promising.

\begin{figure}
   \includegraphics[width=0.50\textwidth]{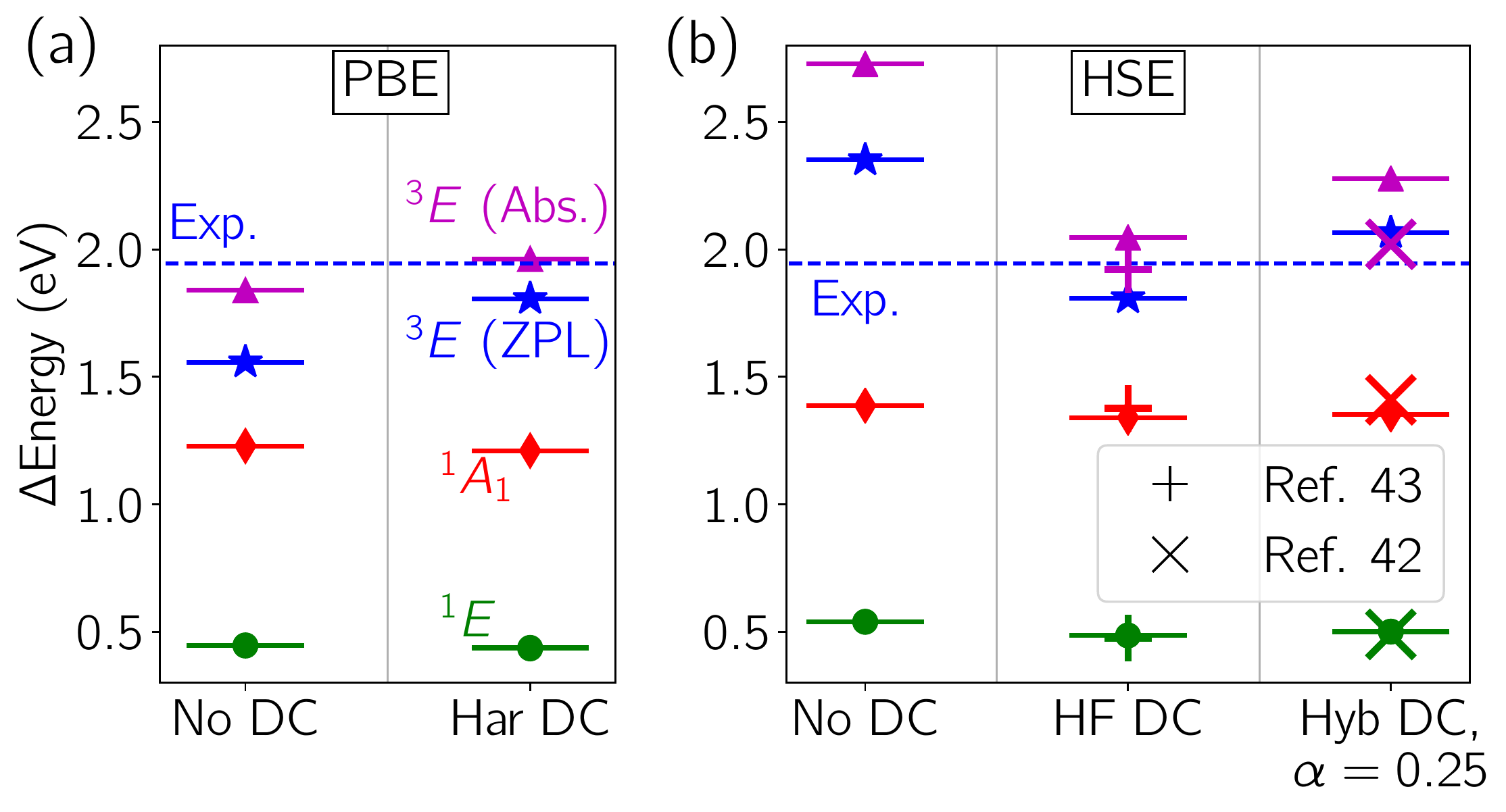}
\caption{\label{fig:NV_DC}
  Energies of many-body excited states of \nv{}, and zero-phonon-line energy for the $^3A_2\rightarrow ^3E$ transition (blue stars, experimental value is blue dashed line) calculated with calculated with either (a) PBE or (b) HSE, with and without double counting (DC) corrections (described in Sec.~\ref{sec:DC}).}
\end{figure}

We see a similar effect for \nv{} in diamond (Fig.~\ref{fig:NV_DC}), where the energies of the MB excited states are increased in the case of the PBE starting point, and decreased for HSE. As with \cbcn{}, the main effect of the DC is to shift the single-particle levels. Thus the $^3E$ energy changes the most, since it involves promotion of an electron from $a_1(2)$ to $e$.  The DC brings the triplet-triplet ZPL (blue stars in Fig.~\ref{fig:NV_DC}) in better agreement with the experimental value (dashed line in Fig.~\ref{fig:NV_DC}).

In addition, we can see that our results match well with the previous implementation of the embedding methods (Ref.~\onlinecite{Ma2020} used the HF DC scheme, and Ref.~\onlinecite{Bockstedte2018} used the hybrid DC), indicating that the general methodology is relatively robust to the details of the calculations, e.g., DFT codes, cRPA implementation, basis, etc. We note that Ma \textit{et al.}~\cite{Ma2020} (who used the HF DC scheme) also used a ``beyond RPA'' strategy which includes the influence of exchange-correlation on the screening, which found a significantly higher value for the $^1A_1$ energy, 1.759 eV compared to 1.376 eV for standard cRPA [standard cRPA is plotted with the red crosses in Fig.~\ref{fig:NV_DC}(b)].

The situation for \feal{} in AlN is significantly more complicated. For \nv{} and \cbcn{}, the DC correction simply renormalized the energy splittings of the excited MB states, via changing the splittings of the single-particle levels. For \feal{}, the nature of the MB ground and low-lying excited states depends sensitively on the splitting of the single particle levels (i.e., the CFS of the Fe $3d$ states, as discussed in Sec.~\ref{sec:feal}). This is known from ligand-field theory, where the $d^5$ Tanabe-Sugano diagram predicts a high-spin $^6A_1$ ground state for small CFS and a low-spin $^2T_2$ for large CFS \cite{Sugano1970}. 

\begin{figure}
   \includegraphics[width=0.50\textwidth]{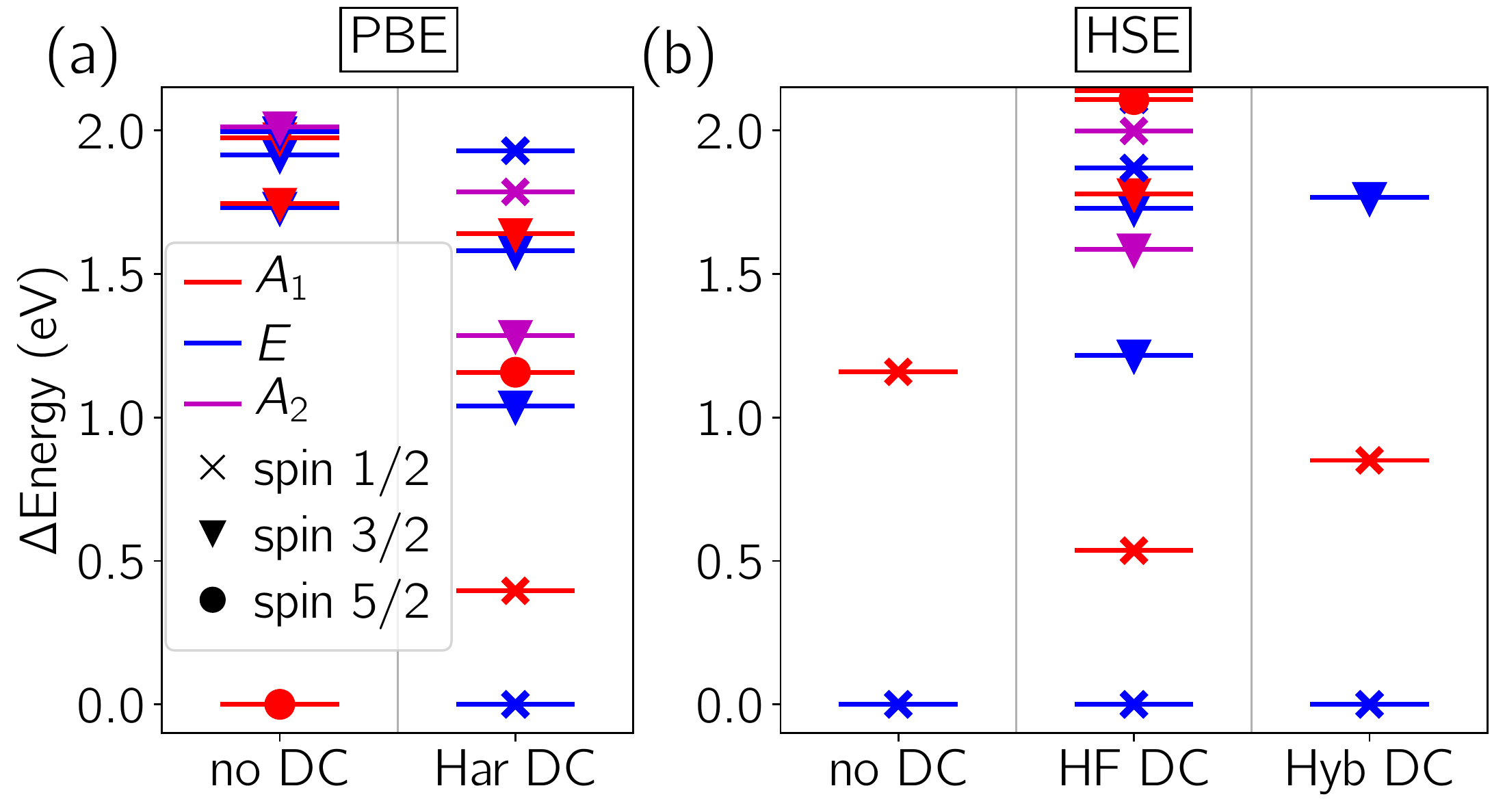}
\caption{\label{fig:feal_DC}
  Energies of many-body ground and excited states of \feal{}, referenced to the lowest energy state, calculated with (a) PBE or (b) HSE, with and without double counting (DC) corrections (described in Sec.~\ref{sec:DC}).
}
\end{figure}

In Fig.~\ref{fig:feal_DC}(a), we show the ground and excited states calculated from a PBE starting point, with and without the Hartree term of the DC [Eq.~(\ref{eq:DC_hartree})]. As discussed above, without DC, the ground state is a 6-fold degenerate spin $5/2$ state with $A_1$ orbital symmetry. The excited states are split from the spin $3/2$ $^4G$ manifold of the free atom by the crystal field. When we add the DC correction, however, the low spin state is favored; the symmetry lowering from the cubic $T_d$ crystal field to the $C_{3v}$ hexagonal one splits the spin $1/2$ $^2T_2$ state into $^2E$ and $^2A_1$ states. For the low-lying excited states, there is a mixture of spin $1/2$ originating from the $^2I$ manifold of the free atom, and the spin $3/2$ states from $^4G$. Also we see the $^6A_1$ state has now moved over 1 eV above the ground state. 

In Fig.~\ref{fig:feal_DC}(b) we plot the energies of the MB states using an HSE starting point. (As discussed above, the occupation was constrained to be the same as the PBE calculation.) We can see that, similar to the PBE calculation with DC correction, the low spin state is the ground state, and the splitting of the $^2T_2$ state from the $C_{3v}$ crystal field is very large. Thus, the other excited states are much higher in energy. Including the HF or hybrid DC correction reduces the splitting somewhat, but it is still much larger than the case of PBE.

As discussed above, it is expected that the ground state should be the high spin $^6A_1$ state. Thus, for PBE, the neglected contribution to the DC from the XC part is clearly critical for obtaining accurate MB energies. In the case of HSE, it seems that the DC is unable to undo the significant overestimation of the CFS, both between the $e$ and $t_2$ manifold from the approximately tetrahedral crystal field, as well as the additional splitting resulting from symmetry-lowering to $C_{3v}$. This could also be caused by neglecting the semilocal XC contribution that should be present in the PBE part of the hybrid. Overall, \feal{} represents a failure of the DC  approaches described in this section, and provides a stringent test case for further development of the DC, and the embedding approach as a whole. 

\section{Discussion \label{sec:disc}}

\subsection{Multireference nature of the many-body states} \label{sec:multiref}

A key utility of embedding methods such as the one described in this work is that they can treat ``multireference'' states, which cannot be described by a single Slater determinant; this goes beyond the capability of, e.g., DFT or Hartree-Fock theory.
Therefore, for understanding the efficacy of the method in going beyond the traditional single-particle theories, it is important to have a metric to understand the degree of multireference nature of the states that we are dealing with. In principle, this can be obtained by analyzing the MB wavefunctions themselves, but care must be taken to differentiate states that are fundamentally multireference, i.e., cannot be expressed as a single Slater determinant in \emph{any basis}, and those that appear multireference because they are a sum of several determinants in our chosen basis.

To accomplish this, we focus on the  one-particle density matrix, $\rho_{i\sigma j\sigma^\prime}=\langle\Psi\vert c^\dagger_{i\sigma}c_{j\sigma^\prime}\vert\Psi\rangle$, where $\Psi$ is a MB state. Then, the MB state can be written as a single Fock state if and only if the density matrix is idempotent, i.e., $\bm{\rho}=\bm{\rho}^2$ \cite{Ballentine2014}. 
To probe this property, we define the quantity 
\begin{equation}
    \Lambda_{\text{MR}}=\text{Tr}(\bm{\rho}-\bm{\rho}^2)=\text{Tr}(\bm{\rho}) - \text{Tr}(\bm{\rho}^2).
    \label{eq:LambdaMR}
\end{equation}
Note that $\Lambda_{\text{MR}}$ is basis-independent due to the cyclic property of the trace. For a state that can be described as a single Slater determinant in some basis, $\Lambda_{\text{MR}}=0$. The maximum value will depend on the number of electrons $N_\text{el}$ available to fill the $2N_\text{orb}$ states (the factor of 2 is for spin). 
Without loss of generality, we choose a basis where $\bm\rho$ is diagonal, so that
\begin{equation}
    \Lambda^{\text{diag}}_{\text{MR}}= \sum_{i=1}^{2N_\text{orb}} \left(  \rho_{ii} - \rho_{ii}^2\right),
    \label{eq:LambdaMRdiag}
\end{equation}
subject to the constraint, $\sum_{i=1}^{2N_\text{orb}} \rho_{ii}= N_\text{el}$.
Then $\Lambda^{\text{diag}}_\text{MR}$ in Eq.~(\ref{eq:LambdaMRdiag}) is maximized when $\rho_{ii} = N_\text{el}/2N_\text{orb}$.
It follows that
\begin{equation}
    \Lambda_\text{MR}^\text{max} = N_\text{el} - 2N_\text{orb}\left( \frac{N_\text{el}}{2N_\text{orb}}\right)^2 = N_\text{el} - \frac{N_\text{el}^2}{2N_\text{orb}}.
\end{equation}
Notice that when $\Lambda_\text{MR}$ reaches its maximum value, $\bm\rho$ is proportional to the identity matrix and thus is basis-independent.

As an example, we consider the MB states of \cbcn{} given in the SM \cite{SM} Table~SI (orbital basis) and Table~SII
(band basis). First we see that, as expected, $\Lambda_{\text{MR}}$ is basis independent. Beginning with the ground-state singlet $\vert\text{GS}\rangle$, we find $\Lambda_{\text{MR}}=0.006$. Since $\vert\text{GS}\rangle$ is a multiorbital singlet that is expected to be multireference \cite{vonBarth1979,Lischner2012}, $\Lambda_{\text{MR}}$ should be finite; however its small value suggests that there is a basis where only one Fock state has the majority of the weight. Indeed, this is the case for the band basis (SM \cite{SM} Table~SII), where the Fock state with two electrons in the bonding state ($\vert 10;10\rangle$) has the vast majority of the weight.
For the triplet $\vert\text{T}\rangle$, the first two MB states (corresponding to $m_s=\pm 1$) are comprised of single Fock states in both bases. Thus $\Lambda_{\text{MR}}=0$, as expected. The third triplet state ($m_s=0$) is ``maximally entangled,'' i.e., $\Lambda_{\text{MR}}=\Lambda^\text{max}_{\text{MR}}=1$. Thus, in both bases, the MB state for $\vert\text{T}\rangle$ with $m_s=0$ involves two Fock states of equal weight. 
The first excited state singlet $\vert\text{D}\rangle$, in contrast with $\vert\text{GS}\rangle$, has $\Lambda_{\text{MR}}=0.991$, close to maximally entangled between two states. In the orbital basis, the state is not qualitatively  distinguishable from $\vert\text{GS}\rangle$, however in the band basis we see that, like $\vert\text{T}\rangle$ with $m_s=0$, the MB state involves a nearly equal superposition of two Fock states. Finally, the second excited singlet $\vert\text{DS}\rangle$ has $\Lambda_{\text{MR}}=0.005$ similar to $\vert\text{GS}\rangle$, consistent with the single Fock state with majority weight in the band basis (SM \cite{SM} Table~SII).
This analysis of \cbcn{} is a clear demonstration of the utility of $\Lambda_{\text{MR}}$. We are able to differentiate between the multideterminant nature of the singlet states, which in certain bases is not \textit{a priori} obvious.

An analysis of $\Lambda_{\text{MR}}$ for \nv{} bears out what is known about the multireference nature of it's MB states. We see in Table~SIII in the SM \cite{SM} that, for the orbital basis, \emph{all} of the states are made up of multiple Fock states. However, for the triplets, there are two states (four for the excited state due to the orbital degeneracy) with $\Lambda_{\text{MR}}=0$, indicating that they \emph{could} be represented by a single Fock state, for a particular choice of basis (as in, e.g., Ref.~\onlinecite{Maze2011}). For the $^3A_2$ state, the band basis in the SM \cite{SM} Table~SIV results in single determinant states. However, in either basis, the excited-state $^3E$ manifold includes mixtures of different Fock states. 
The singlets all have  $\Lambda_{\text{MR}}\simeq1$. 
This does not represent maximal entanglement, since for \nv{} we have eight total spin-orbitals and six electrons so $\Lambda^{\text{max}}_{\text{MR}}=1.5$. Thus the singlets and $m_s=0$ triplet states are not maximally entangled in the context of our full Hilbert space. 
However, they are maximally entangled with respect to a smaller Hilbert space; the \nv{} MB states primarily involve Fock states with two of the orbitals completely filled by two electrons, and the other two half filled with one spin. This is most clearly seen in the $m_s=0$ state of $^3A_2$ in the band basis (second row of Table~SIV in the SM \cite{SM}, where the Fock states have fully occupied $a_1(1)$ and $a_1(2)$ states, and half occupied $e$ states. Thus the entanglement occurs between four spin-orbitals occupied by two electrons; analogous to the case of \cbcn{},  $\Lambda^{\text{max}}_{\text{MR}}$ for this reduced space is unity.

For the case of \feal{}, the only single-determinant states we find is the $m_s=\pm5/2$ states of the high-spin $^6A_1$ manifold. All other states have $\Lambda_{\text{MR}}>1$, and many of them  (SM \cite{SM} Table~SV) are close to maximally entangled, i.e., $\Lambda_{\text{MR}} \simeq \Lambda^{\text{max}}_{\text{MR}}= 2.5$. 

\subsection{Obtaining simplified models for defect interactions \label{sec:models}}

In general, the Coulomb interaction is represented by the full four-index $U_{ijkl}$ tensor. Given the minimal basis sets we use for the description of the embedded correlated states, we aim in the following to minimize the number of needed parameters \emph{even further}, by neglecting more and more channels of the full Coulomb interaction tensor.
Indeed, in an atomic-orbital picture, the type of orbitals and the point symmetry govern which elements of $U_{ijkl}$ are present; thus by comparing the results from our Wannier basis with those expected assuming atomic-like orbitals, we can gain insight into the interaction between the defect states, and with the bulk.
Also, simplified interactions are useful for creating minimal models for further analysis, and may be required for the use of some MB solvers. 

We will consider reducing $U_{ijkl}$ to a two-index tensor, where $U_{ij}$ includes the intra- and inter-orbital density-density interactions ($k=i$ and $l=j$) and $J_{ij}$ includes the Hunds couplings ($i=l$ and $k=j$). Also, we will consider using just three average parameters: $U=\frac{1}{N_{\text{orb}}}\sum_i U_{iiii}$
$U^\prime=\frac{1}{N_{\text{orb}}(N_{\text{orb}}-1)}\sum_{i\ne j} U_{ijij}$, and $J=\frac{1}{N_{\text{orb}}(N_{\text{orb}}-1)}\sum_{i\ne j} U_{ijji}$ to construct the interaction. We will keep the full noninteracting part of Eq.~(\ref{eq:MB}) from our Wannier calculations; in this section, we will focus on PBE calculations and, to simplify the discussion, neglect the DC correction. 

\begin{figure}
   \includegraphics[width=\columnwidth]{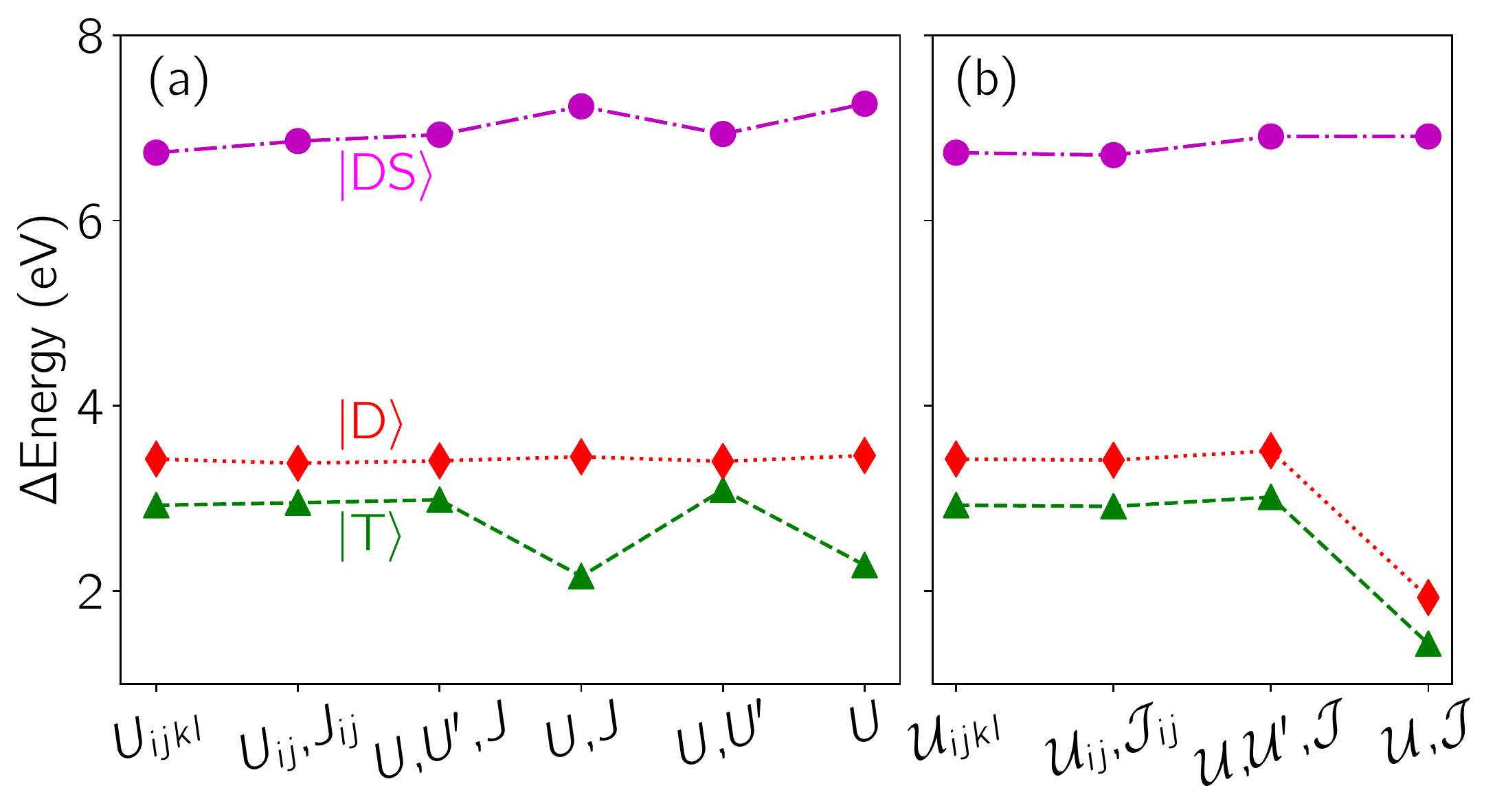}
\caption{\label{cbcn_U} Energy of the many-body states of \cbcn{} in BN, calculated with PBE and no double-counting correction, with respect to the ground-state singlet state for different simplified interactions (see Sec.~\ref{sec:models}), where orbital averaging is performed using the interaction in the (a) orbital, or (b) band basis. 
The $x$ labels denote: $U_{ijkl}$: full interaction tensor;  $U_{ij},J_{ij}$: two component interactions; $U,U^\prime,J$: orbitally averaged intraorbital and interorbital density-density and Hunds exchange interactions; $U,J$: orbitally averaged intraorbital density-density and  Hunds interactions; $U,U^\prime$: orbitally averaged intraorbital and interorbital density-density interactions; $U$: orbitally averaged intraorbital interactions. Script versions in (b) correspond to the same quantities, but averaged in the band basis.}
\end{figure}

In Fig.~\ref{cbcn_U} we plot the energies of the MB states with respect to $\vert\text{GS}\rangle$ using these simplified screened Coulomb interaction tensors. Recall that when we solve the Hamiltonian in Eq.~(\ref{eq:MB}) including all terms in $U_{ijkl}$, the orbital and band bases are related by a unitary transformation, and thus result in the same MB energies; however, once we start simplifying the interaction (i.e., removing or averaging terms), the resulting MB energies will depend on which basis for $U_{ijkl}$ that we start from. Thus we plot the energies with simplified interactions starting from both the orbital [Fig.~\ref{cbcn_U}(a)] and band [Fig.~\ref{cbcn_U}(b)] bases (script letters will denote averaging in the band basis). 

As discussed in Sec.~S2 A 1 of the SM \cite{SM}, the energies of the states of \cbcn{} can be estimated with just one interaction parameter $U$, though we construct this value from the difference between the intraorbital and interorbital density-density interaction terms.
Simplifying $U_{ijkl}$ to a two-component form does not result in a significant change in the MB spectrum for either basis; e.g., for the orbital basis in Fig.~\ref{cbcn_U}(a) all the excited states shifted up in energy by less than 125 meV. Performing an average over the orbitals to obtain effective parameters, we find $U=1.94$ eV, $U^\prime=1.41$ eV, and $J=0.08$ eV for the orbital basis. Taking these orbitally averaged values as the Coulomb interaction only results in changes in the energy at the meV level. Neglecting the Hunds $J$ has a minor effect on the splitting of the spin states, shifting the triplet upward in energy (i.e., closer to the corresponding singlet $\vert \text{D}\rangle$) by 166 meV. Both the intraorbital ($U$) and interorbital ($U^\prime$) terms are necessary for obtaining accurate energies for the triplet state [see Fig.~\ref{cbcn_U}(a)]. 

Performing the averaging in the band basis, we obtain $\mathcal{U}=1.77$ eV, $\mathcal{U}^\prime=1.58$ eV, and $\mathcal{J} = 0.25$ eV. The significantly larger value of $\mathcal{J}$ compared to the orbital basis is indicative of the importance of exchange in this basis: neglecting $\mathcal{J}$ does not produce the correct spin states [and thus these points are not included in Fig.~\ref{cbcn_U}(b)]. Specifically, $\mathcal{J}$ is necessary to capture the $m_s=0$ triplet state (otherwise it becomes a spin $1/2$ doublet). In the band basis we also see in Fig.~\ref{cbcn_U}(b) that both intraorbital and interorbital density-density interactions are required to accurately capture the energies of $\vert\text{T}\rangle$ and $\vert\text{D}\rangle$.

\begin{figure}
   \includegraphics[width=\columnwidth]{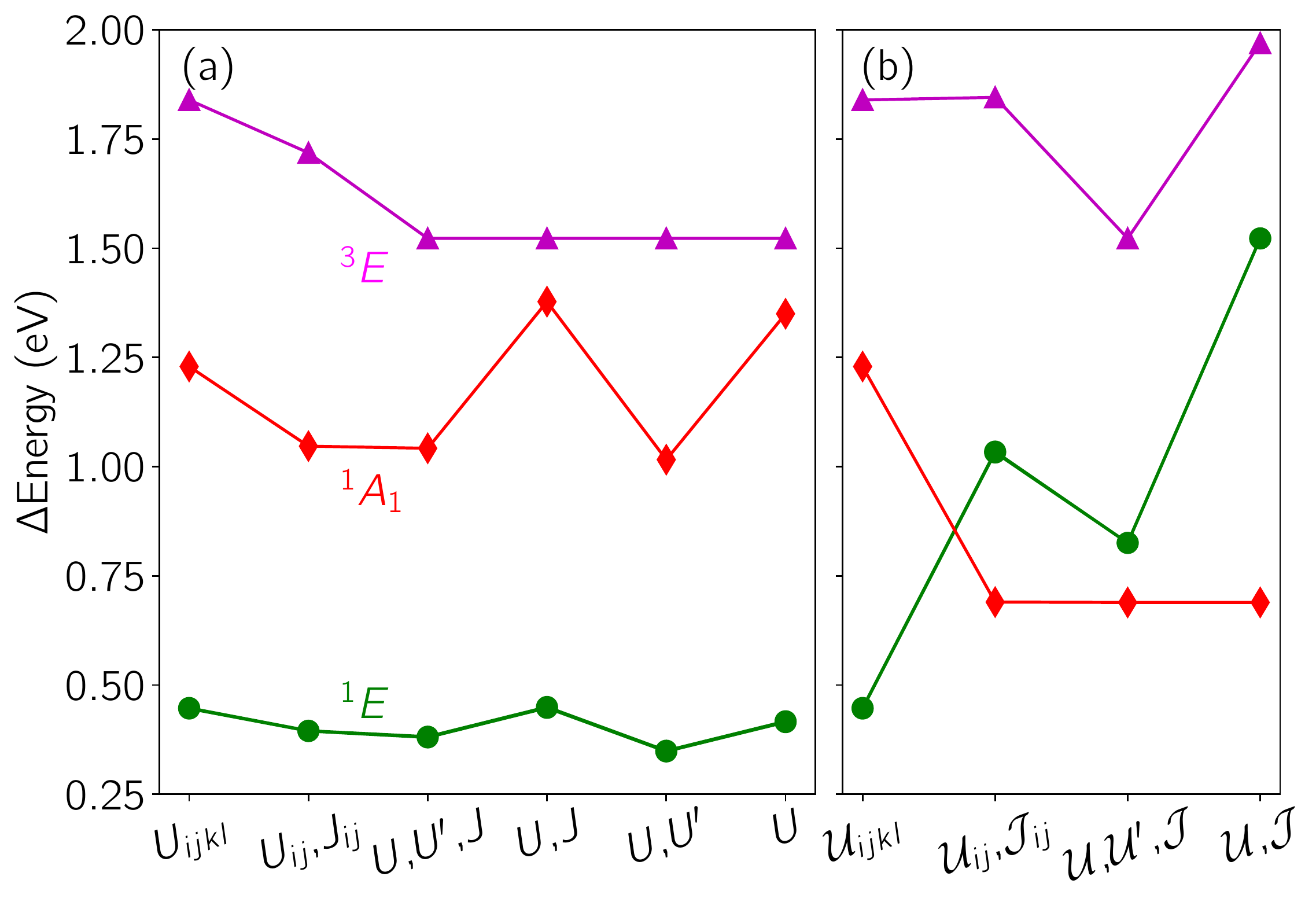}
\caption{\label{NVU} Energies of the many-body states of \nv{} in diamond calculated using the PBE functional (no double-counting corrections), with respect the ground-state triplet for different simplified interactions (see Sec.~\ref{sec:models}), where orbital averaging is performed using the interaction in the (a) orbital, or (b) band basis. The $x$ labels denote: $U_{ijkl}$: full interaction tensor;  $U_{ij},J_{ij}$: two component interactions; $U,U^\prime,J$: orbitally averaged intraorbital and interorbital density-density and Hunds exchange interactions; $U,J$: orbitally averaged intraorbital density-density and  Hunds interactions; $U,U^\prime$: orbitally averaged intraorbital and interorbital density-density interactions; $U$: orbitally averaged intraorbital interactions. Script versions in (b) correspond to the same quantities, but averaged in the band basis.}
\end{figure}

Though the \nv{} center has a more complex electronic structure than \cbcn{}, we can still hope to gain insight into the MB states and the possibility for simplified models via exploring reduced forms of the Coulomb interaction. In Fig.~\ref{NVU}, we plot the energies of the MB states with such simplifications, either based on the $U_{ijkl}$ tensor in the orbital [Fig.~\ref{NVU}(a)], or the band [Fig.~\ref{NVU}(b)] basis. 

Similar to \cbcn{}, the case of the orbital basis is better behaved for increasingly simple descriptions of the Coulomb interaction. In this basis, the $^3A_2-^3E$ triplet-triplet splitting decreases slightly when reducing to a two-coordinate form of the interaction, and then further when an orbitally-averaged interaction is used (``$U,U^\prime,J$'' in Fig.~\ref{NVU}(a), where $U=2.43$ eV, $U^\prime=0.81$ eV, and $J=0.03$ eV for the orbital basis). Further simplification of the Coulomb interaction does not change the triplet-triplet splitting.
This behavior is likely because the $e$ states are equal superpositions of the dangling bonds on the C atoms around the vacancy \cite{Maze2011}, so the excited state triplet more or less involves equal population of these states. Interestingly, the first excited state singlet $^1E$ energy in Fig.~\ref{NVU}(a) is also only mildly effected by the treatment of the interaction, while the energy splitting to the $^1A_1$ state is the most sensitive to the treatment of interactions, especially whether or not the inter-orbital interaction $U^\prime$ is included.

The case of the band basis [Fig.~\ref{NVU}(b)] is similar to that of \cbcn{} in that Hunds $\mathcal{J}$ is necessary for a description of the entangled $m_s=0$ triplet states, and the band basis results in a smaller effective onsite $\mathcal{U}=1.57$ eV, and larger $\mathcal{U}^\prime=1.096$ eV and $\mathcal{J} =0.31$ eV than for the orbital basis. Also analogously to \cbcn{}, the MR nature of the states are significantly reduced for simplified Coulomb interactions. Most strikingly, we see in Fig.~\ref{NVU}(b) that systematic simplification of the interaction, even down to the two-index tensor, produces an incorrect energetic ordering; this is in contrast to the Wannier-orbital basis  [Fig.~\ref{NVU}(a)], where the qualitative properties of the MB states are correct even if we use only a single interaction parameter.

\subsection{Summary and implications from test cases}

The aim of this work was to critically review the embedding approach for describing the excited states of point defects. What is clear from the results in Sec.~\ref{sec:lessons} is that this method can qualitatively treat a variety of different types of excited states. In all cases, numerical convergence was fairly straightforward. The bulk screening in the cRPA could be converged via increasing the number of bands per atom, as well as the $k$ mesh density and/or the supercell size. The MB energies were also relatively insensitive to the details of the Wannierization procedure. Regarding the choice of XC functional, it was clear that the main difference between HSE and PBE is the increased splitting in the single particle states in the hybrid functional. Even so, using the appropriate DC correction for PBE and HSE resulted in significantly improved agreement between the two starting points. For \cbcn{} and \nv{}, the resulting MB energies were also in good agreement with available experimental observations. For \nv{}, our calculations were consistent with previous embedding implementations \cite{Bockstedte2018,Ma2020}.

\feal{} in AlN constituted the most challenging case, where we showed that the nature of the MB ground and excited states was extremely sensitive to the magnitude of the CFS, which itself was very sensitive to the XC potential of the initial DFT calculation. In this case the DC correction was not sufficient to reproduce the expected high-spin ground state with either PBE and HSE. We expect that including the PBE XC contribution to the DC will perhaps alleviate this issue. It is also possible that a more accurate treatment of the bulk screening beyond the RPA approximation is necessary, as proposed in Ref.~\onlinecite{Galli2021}.
In any case, \feal{} constitutes a significant challenge to the embedding methodology, and thus an excellent test case for future developments.

\section{Conclusions} \label{sec:conclusions}

In this work, we critically reviewed an embedding approach to treat correlated excited states of point defects. The method is based on Wannierization of density-functional theory calculations in order to obtain an active space, including Coulomb interactions in that active space via the constrained RPA method, and correcting for the interaction already included in the DFT part with a functional-dependent double-counting scheme. We showed that this approach provides quantitative accuracy for the \cbcn{} defect in BN and \nv{} in diamond, though the more complex and sensitive electronic structure of \feal{} in AlN represents a challenge that is out of reach of the present methodology.  
Overall, despite the complexity and yet unanswered questions about the methodology, we conclude that quantum embedding represents a promising approach to describing the correlated excited states of a variety of point defects in materials.

\acknowledgements

CED thanks A. Alkauskas, D. Wickramaratne, M. Zingl, A. Gali, M. Turiansky, T. Berkelbach, and A. Millis for fruitful conversations and comments on the manuscript. The Flatiron Institute is a division of the Simons Foundation. CED acknowledges support from the National Science Foundation under Grant No. DMR-1918455. The work of DIB was supported by the grant of the President of the Russian Federation, Project SP-2488.2021.1.

\bibliography{defects}

\end{document}


\title{Supplemental Material: Quantum embedding methods for correlated excited states of point defects: \\Case studies and challenges}
\author{Lukas Muechler}
\affiliation{Center for Computational Quantum Physics, Flatiron Institute, 162 5$^{th}$ Avenue, New York, NY 10010}
\author{Danis I. Badrtdinov}
\affiliation{Theoretical Physics and Applied Mathematics Department,
Ural Federal University, 620002 Yekaterinburg, Russia}
\affiliation{Center for Computational Quantum Physics, Flatiron Institute, 162 5$^{th}$ Avenue, New York, NY 10010}
\author{Alexander Hampel}
\affiliation{Center for Computational Quantum Physics, Flatiron Institute, 162 5$^{th}$ Avenue, New York, NY 10010}
\author{Jennifer Cano}
\affiliation{Center for Computational Quantum Physics, Flatiron Institute, 162 5$^{th}$ Avenue, New York, NY 10010}
\affiliation{Department of Physics and Astronomy, Stony Brook University, Stony Brook, New York 11794-3800, USA}
\author{Malte R\"osner}
\affiliation{Radboud University, Institute for Molecules and Materials, Heijendaalseweg 135, 6525 AJ Nijmegen, Netherlands}
\author{Cyrus E. Dreyer}
\affiliation{Center for Computational Quantum Physics, Flatiron Institute, 162 5$^{th}$ Avenue, New York, NY 10010}
\affiliation{Department of Physics and Astronomy, Stony Brook University, Stony Brook, New York 11794-3800, USA}
\date{\today}

\maketitle

\section{Computational parameters \label{sec:comp_param}}

\subsection{\cbcn{} in hexagonal BN} \label{sec:compBN}

In this work we will be studying the neutral charge state of \cbcn{} in bulk layered hexagonal BN ($P6_3/mmc$), which involves replacing a nearest-neighbor B and N pair with a C dimer. The basic electronic structure is discussed in the main text Sec.~II~A. As with any treatment of point defects using periodic boundary conditions, the supercell size must be converged so that the defect is isolated from its periodic images \cite{Freysoldt2014}. Since here we are treating a neutral defect, the electrostatic interaction between defect wavefunctions is not an issue. Structurally, the bonds around the carbon dimer are well converged for a $5\times5\times1$ supercell (100 atoms): the C-C, C-N, and C-B bonds change less than $5\times10^{-4}$\AA{} when the cell is increased to $6\times6\times1$. In the main text Sec.~IV~D~2, we discuss the convergence necessary to obtain accurate RPA bulk screening of the defect subspace, demonstrating that, for a $5\times5\times1$ supercell that the many-body (MB) energies converge for around 8 bands per atom, and convergence can be improved with respect to Brillouin-zone sampling by increasing the $k$ mesh in addition to supercell size. We note that this can be achieved by increasing the $k$ mesh just in plane, as the hopping out of plane is quite small: 0.09 eV, which can be reduced to 3 meV by doubling the cell in the $c$ direction. However, we find that using a larger cell in $\hat{c}$ does not significantly change the many-body energies ($<20$ meV).

Thus, for the PBE \cite{Perdew1996} and HSE \cite{HSE2003,HSE2006} calculations in the main text, we will use results from the $5\times5\times1$ supercell, with a $4\times4\times2$ $\Gamma$-centered (to preserve the hexagonal symmetry) $k$ mesh. All calculations are performed with an energy cutoff of 500 eV, which was demonstrated to provide accurate electronic structure in, e.g., Refs.~\onlinecite{Maciaszek2022,Mackoit2019,Dreyer2014}. The same standard VASP PAW potentials were used as in Ref.~\onlinecite{Dreyer2014}.

\subsection{\nv{} in diamond}

The negatively-charged nitrogen vacancy center in diamond (\nv{}) is constructed by creating a vacancy in diamond, and replacing a nearest-neighbor C with N. This defect is charged, so in principle charge-state corrections (e.g., Ref.~\onlinecite{Freysoldt2009}) are necessary to determine, e.g., charge-state transition levels \cite{Freysoldt2014}. However, in our case, we are only interested in neutral excitations; since we are not changing the charge state, these corrections are not necessary. Indeed, we see rapid convergence of the MB energies with supercell size (see Fig.~5 in the main text).
Thus, for the calculations in the main text we use 215 atom supercells, one $k$-point at $\Gamma$, and an energy cutoff of 500 eV. Such energy cutoff and standard VASP PAW potentials have been shown previously to provide an excellent description of the properties of \nv{} (e.g., in Refs.~\onlinecite{Alkauskas2014,Razinkovas2021}).

\subsection{\feal{} in AlN \label{sec:feal_conv}}

For the Fe substituting Al in AlN (\feal{}), we also consider the neutral charge state. As we showed in the main text (see Fig.~8), convergence can be accelerated in the same way as for \cbcn{}, i.e., increasing supercell size and $k$ mesh density at the same time. 
For the results in the main text, we will use a 128 atom supercell, $4\times4\times4$ $\Gamma$-centered $k$-mesh, and 500 eV cutoff. The cutoff is in excess of what was demonstrated to provide accurate structural properties and electronic structure for AlN and \feal{} in Refs.~\onlinecite{Wickramaratne2016,Wickramaratne2019}, and the standard VASP PAW potentials used here are the same as those in that study.

\section{Details of the many-body states \label{sec:mb_details}}

In this section we provide details of the MB wavefunctions that we calculate using the methodology in Sec.~III. 
In general, we consider two bases, referred to in the main text as ``orbital'' and ``band.'' The orbital basis is the occupation basis of the Wannier functions used in the calculation, whether they are projected or localized. The band basis is the one that diagonalizes $t_{ij}$. For case where no disentanglement is used, the eigenvalues of $t_{ij}$ correspond to the energies of the Kohn-Sham states. For simplicity,m, the PBE functional and no double-counting correction is used for the results in this section.

\subsection{\cbcn{} in BN \label{sec:cbcn_mb}}

\begin{table*}
  \caption{\label{tab:cbcn_param} Wavefunction parameters, energies (with respect to the ground state singlet, based on PBE DFT calculations), spin moments, and multireference character of the states of \cbcn{} in the orbital basis ($\vert \overline{p^2_z} \overline{p^1_z}; p^2_z p^1_z\rangle$). }
\begin{ruledtabular}
  \begin{tabular}{c|cccc}
   &  $\Delta E$ (eV) & Many-body state & $S,m_s$ & $\Lambda_\text{MR}$  \\
    \hline
    $\vert \text{GS}\rangle$& --  & $0.161\vert 01;01\rangle  + 0.409\vert 01;10\rangle  + 0.409\vert 10;01\rangle  + 0.799\vert 10;10\rangle $ &0, 0&  0.006\\\hline
    \multirow{ 3}{*}{$\vert \text{T}\rangle$}  & \multirow{ 3}{*}{2.926}& $\vert 00;11\rangle $ &1, 1& 0.0\\
     & & $\vert 11;00\rangle $ &1, -1&0.0\\
     & &$-0.707\vert 01;10\rangle  + 0.707\vert 10;01\rangle $ &1, 0&1.0\\\hline
    $\vert \text{D}\rangle$ & 3.425 &$0.502\vert 01;01\rangle  + 0.458\vert 01;10\rangle  + 0.458\vert 10;01\rangle  - 0.571\vert 10;10\rangle $&0, 0&0.991\\\hline
    $\vert \text{DS}\rangle$ & 6.732& $-0.849\vert 01;01\rangle  + 0.349\vert 01;10\rangle  + 0.349\vert 10;01\rangle  - 0.186\vert 10;10\rangle $ &0, 0& 0.005\\
\end{tabular}
\end{ruledtabular}
\end{table*}

\begin{table*}
  \caption{\label{tab:cbcn_param_band} Wavefunction parameters, energies (with respect to the ground state singlet, based on PBE DFT calculations), spin moments, and multireference character of the states of \cbcn{} in the band basis ($\vert \overline{b_2} \overline{b^*_2}; b_2 b^*_2\rangle$). }
\begin{ruledtabular}
  \begin{tabular}{c|cccc}
   & $\Delta E$ (eV)& Many-body state  & $S,m_s$ & $\Lambda_\text{MR}$  \\
    \hline
    $\vert \text{GS}\rangle$  & -- & $-0.038\vert01;01\rangle + 0.022\vert01;10\rangle + 0.022\vert10;01\rangle + 0.999\vert10;10\rangle$ &0, 0&  0.006\\\hline
    \multirow{ 3}{*}{$\vert \text{T}\rangle$}  &  \multirow{ 3}{*}{2.926} &$\vert 00;11\rangle $&  1, 1& 0.0\\
     & &$\vert 11;00\rangle $ &1, -1&0.0\\
     & &$-0.707\vert 01;10\rangle  + 0.707\vert 10;01\rangle $&1, 0&1.0\\\hline
    $\vert \text{D}\rangle$ & 3.425& $0.035\vert01;01\rangle - 0.706\vert01;10\rangle - 0.706\vert10;01\rangle + 0.033\vert10;10\rangle$&0, 0&0.991\\\hline
    $\vert \text{DS}\rangle$ & 6.732& $-0.999\vert01;01\rangle - 0.026\vert01;10\rangle - 0.026\vert10;01\rangle - 0.037\vert10;10\rangle$ &0, 0& 0.005\\
\end{tabular}
\end{ruledtabular}
\end{table*}

In Table \ref{tab:cbcn_param}, we give the MB wavefunctions, energies (with respect to the ground-state singlet), and $\Lambda_{\text{MR}}$ for \cbcn{} in the orbital basis (calculations are performed with the PBE functional). The Fock states are labeled as $\vert \overline{p^2_z} \overline{p^1_z}; p^2_z p^1_z\rangle$, where 1 and 2 label $p_z$ orbitals on the two C atoms. Table \ref{tab:cbcn_param_band} presents the same states, but in the band basis, $\vert \overline{b_2} \overline{b^*_2}; b_2 b^*_2\rangle$ where $b_2$ is the bonding combination of C $p_z$ orbitals, and $b^*_2$ is the antibonding combination. We see that, the specific weights of the Fock states of the singlets are basis dependent. We also see that, as expected, $\Lambda_{\text{MR}}$ and the energies are basis independent.

\subsubsection{Comparison with Hubbard dimer} \label{sec:cbcn_mb_hd}

We can understand the nature of the MB states of \cbcn{} and their energetic ordering by comparison to a simple Hubbard model of a two-orbital dimer molecule. The analysis in this section will follow Refs.~\onlinecite{Ocampo2017} and \onlinecite{Carrascal2015}.
In analogy to \cbcn{}, we denote the orbitals on the dimer atoms as $p_z^1$ and $p_z^2$, so our basis is $\vert \overline{p^2_z} \overline{p^1_z}; p^2_z p^1_z\rangle$ (recall, the overbar indicates spin down). We consider a simplified Hamiltonian with two parameters: an onsite Hubbard $U$ and intersite hopping $t$ (we assume no energy splitting between the levels). At half filling, we have the Hamiltonian \cite{Carrascal2015,Ocampo2017}
\begin{equation}
\label{Hdimer}
\begin{split}
H_{\text{dimer}}&=-\frac{U}{2}\left[(n_{p_z^1\uparrow}-n_{p_z^1\downarrow})^2+(n_{p_z^2\uparrow}-n_{p_z^2\downarrow})^2\right] 
\\
&+t\sum_\sigma (c^\dagger_{p_z^1\sigma}c_{p_z^2\sigma}+\text{H.c.}),
\end{split}
\end{equation}
where $t$ is the hybridization between the orbitals and $U$ is the onsite Coulomb repulsion. We assume the single-particle states of the dimer are degenerate. The ground state of this model is a spin singlet with the wavefunction
\begin{equation}
\begin{split}
    \vert \text{GS}\rangle & = \frac{1}{a}\Bigg[\vert 10;01 \rangle + \vert 01;10 \rangle 
    \\
    &- \frac{4t}{U+\sqrt{U^2+16t^2}}\left(\vert 10;10 \rangle+\vert 01;01 \rangle \right)\Bigg],
    \end{split}     
\end{equation}
where $a$ is the normalization constant. We can see that for the noninteracting case, the ground state is has equal weight on all single-particle states. For $U \gg t$, there is no double occupation of orbitals. We see from Table \ref{tab:cbcn_param} that the coefficients of the states $\vert 10;10\rangle$ and $\vert 01;01\rangle$ are different for \cbcn{}, which is a result of the fact that (unlike in the dimer model) the onsite hoppings on the two C sites differ (one C is bonded to two N atoms, and the other C to two B atoms).

The dependence on $U/t$ is more clear if we were to change to this basis ($\vert \overline{b_2} \overline{b^*_2}; b_2 b^*_2\rangle$) of hybridized orbitals. The ground state would have the relative occupation of the antibonding state to the bonding state to be $U/(4t-\sqrt{U^2+16t^2})$. Therefore, the noninteracting state would be double occupation of the bonding orbital, and $U$ mixes in a contribution from the antibonding state. We see in the first row in Table \ref{tab:cbcn_param_band} that the MB state for \cbcn{} with this basis has majority weight on the Fock state $\vert10;10\rangle$, corresponding to two electrons in the bonding state. In addition to $\vert01;01\rangle$ predicted from the dimer model,  there are small contributions from the ``mixed'' singlet due to the symmetry breaking between sites.
Using effective values for or $U$ and $t$ derived from our \textit{ab initio} calculations (see Sec.~V~B in the main text), the weight of the ``antibonding'' single particle state to the GS wavefunction is $U/(4t-\sqrt{U^2+16t^2})=-0.10$, which is a factor of two larger than the weight $\vert01;01\rangle$ Fock state in Table \ref{tab:cbcn_param_band}.

The ground state energy of this model is  $E_{\text{GS}}=\frac{U}{2}-\frac{1}{2}\sqrt{U^2+16t^2}$; the next state in the spectrum is a triplet with energy $E_{\text{T}}=-U$; finally, there are two excited-state singlets with energy $E_{\text{D}}=0$ and $E_{\text{DS}}=\frac{U}{2}+\frac{1}{2}\sqrt{U^2+16t^2}$ \cite{Ocampo2017,Carrascal2015}.
Using the model parameters (in terms of our calculated  $t_{ij}$ and $U_{ijkl}$), $U=(U_{p^1_zp^1_zp^1_zp^1_z}+U_{p^2_zp^2_zp^2_zp^2_z})/2-U_{p^1_zp^1_zp^2_zp^2_z}=0.53$ eV, where we subtract the nearest-neighbor Coulomb interaction from the local interactions (similar to the $U^*$ approach from Ref.~\onlinecite{schuler_optimal_2013}), and $t=t_{p^1_zp^2_z}=-1.32$, eV we find energy splittings of $E_{\text{T}}-E_{\text{GS}}=1.86$ eV, $E_{\text{D}}-E_{\text{GS}}=2.39$ eV, and $E_{\text{DS}}-E_{\text{GS}}=5.31$ eV, in quite good qualitative agreement with the calculated values in Table \ref{tab:cbcn_param}. Quantitative agreement can be improved by also including the calculated energy splitting between the orbitals, i.e., the exact noninteracting part from our DFT calculation, but a single $U$ parameter.

\subsubsection{Double-Counting correction in \cbcn{} from the dimer model \label{sec:dimer_dc}}

 Based on the quantitative similarity between the dimer model and the MB spectrum of  \cbcn{}, we can use the results of Refs.~\onlinecite{vanLoon2021} to derive a double-counting (DC) correction. In that work, the authors compared the dimer model solved with KS-DFT to exact calculations. They demonstrated that the energies for the bonding and antibonding states in the Hubbard Dimer model can be written as (in our notation) $\epsilon^{\text{DFT}}_{b_2}=\epsilon_{b_2}+U_{b_2b_2b_2b_2}$ and $\epsilon^{\text{DFT}}_{b^*_2}=\epsilon_{b^*_2}+U_{b^*_2b_2b^*_2b_2}$, where $\epsilon$ is the noninteracting energy and $\epsilon^{\text{DFT}}$ includes DFT Coulomb interaction using the \emph{exact} exchange-correlation (XC) functional. These expressions apply for situations where the onsite interaction is much larger than the Hunds coupling ($U/J > 5$ was demonstrated in Ref.~\onlinecite{vanLoon2021}), as is the case for \cbcn{} ($U_{b_2b_2b_2b_2}/U_{b^*_2b_2b_2b^*_2}=6.7$ using PBE numbers). Thus, the DC correction will shift the single-particle states like $\Delta\epsilon^{\text{DFT}}=U_{b_2b_2b_2b_2}-U_{b^*_2b_2b^*_2b_2}$. Since the intraorbital interaction is slightly larger than the interorbital one, the DC correction slightly increases the splitting between the single particle states, and thus the MB excited-state energies (see ``Dimer DC'' in Figs.~9(a) and 10(a) in the main text).

\subsection{\nv{} in diamond}

In Table \ref{tab:NV_param} we give the details of the MB wavefunctions for \nv{} (PBE functional) in the orbital basis. This basis consists of three $sp^3$ dangling bonds from C atoms around the vacancy (denoted $sp^3_{\text{C}_i}$, $i=1,2,3$) and the N atom ($sp^3_{\text{N}}$): $\vert \overline{sp^3_{\text{C}_1}}\overline{sp^3_{\text{C}_2}}\overline{sp^3_{\text{C}_3}}\overline{sp^3_{\text{N}}}; sp^3_{\text{C}_1}sp^3_{\text{C}_2}sp^3_{\text{C}_3}sp^3_{\text{N}}\rangle$. In Table~\ref{tab:NV_param_band}, we show the same states in the band basis, $\vert \overline{e_xe_ya_1(2)a_1(1)} ; e_xe_ya_1(2)a_1(1)\rangle$. As with \cbcn{}, $\Lambda_{\text{MR}}$ and the energies are basis independent as expected.

\begin{table*}
  \caption{\label{tab:NV_param} Wavefunction parameters, energies (with respect to the ground state triplet, based on PBE DFT calculations), spin moments, and multireference character of the states of \nv{}  in the orbital basis ($\vert \overline{sp^3_{\text{C}_1}}\overline{sp^3_{\text{C}_2}}\overline{sp^3_{\text{C}_3}}\overline{sp^3_{\text{N}}}; sp^3_{\text{C}_1}sp^3_{\text{C}_2}sp^3_{\text{C}_3}sp^3_{\text{N}}\rangle$).}
  \centering
\begin{ruledtabular}
  \begin{tabular}{c|c|>{\centering\arraybackslash}p{13cm}cc}
   & $\Delta E$ (eV) & Many-body state(s) & $S,m_s$ &$\Lambda_\text{MR}$  \\
    \hline
    \multirow{ 3}{*}{$^3A_2$} & \multirow{ 3}{*}{--}& $0.577(\vert1111;0011\rangle + \vert1111;0101\rangle + \vert1111;1001\rangle)$ &1,-1& 0.0\\\cline{3-5}
    & &$0.408(\vert1011;0111\rangle-\vert0111;1011\rangle - \vert0111;1101\rangle - \vert1011;1101\rangle + \vert1101;0111\rangle + \vert1101;1011\rangle)$ &1,0& 1.0\\\cline{3-5}
    & & $0.577(\vert0011;1111\rangle +\vert0101;1111\rangle + \vert1001;1111\rangle)$ &1,1& 0.0\\\hline
    \multirow{ 2}{*}{$^1E$}  &   \multirow{ 2}{*}{0.428} &$0.056\vert0111;0111\rangle + 0.518\vert0111;1011\rangle + 0.401\vert0111;1101\rangle - 0.013\vert0111;1110\rangle + 0.518\vert1011;0111\rangle + 0.194\vert1011;1011\rangle - 0.116\vert1011;1101\rangle + 0.046\vert1011;1110\rangle + 0.4010\vert1101;0111\rangle - 0.116\vert1101;1011\rangle -0.250\vert1101;1101\rangle + 0.060\vert1101;1110\rangle - 0.013\vert1110;0111\rangle + 0.046\vert1110;1011\rangle + 0.059\vert1110;1101\rangle$ &0,0&1.0\\\cline{3-5}
    &&$ -0.257\vert0111;0111\rangle - 0.164\vert0111;1011\rangle + 0.366\vert0111;1101\rangle + 0.061\vert0111;1110\rangle - 0.164\vert1011;0111\rangle + 0.177\vert1011;1011\rangle + 0.53\vert1011;1101\rangle + 0.042\vert1011;1110\rangle + 0.366\vert1101;0111\rangle + 0.53\vert1101;1011\rangle + 0.080\vert1101;1101\rangle - 0.019\vert1101;1110\rangle + 0.061\vert1110;0111\rangle + 0.042\vert1110;1011\rangle - 0.019\vert1110;1101\rangle$ &0,0&1.0\\\hline
    $^1A_1$ & 1.198 & $-0.392\vert0111;0111\rangle - 0.298\vert0111;1011\rangle + 0.298\vert0111;1101\rangle + 0.03\vert0111;1110\rangle - 0.298\vert1011;0111\rangle - 0.392\vert1011;1011\rangle - 0.298\vert1011;1101\rangle - 0.03\vert1011;1110\rangle + 0.298\vert1101;0111\rangle - 0.298\vert1101;1011\rangle - 0.393\vert1101;1101\rangle + 0.030\vert1101;1110\rangle + 0.030\vert1110;0111\rangle - 0.030\vert1110;1011\rangle + 0.030\vert1110;1101\rangle + 0.015\vert1110;1110\rangle$& 0,0&1.09\\\hline
    \multirow{ 6}{*}{$^3E$} &\multirow{ 6}{*}{1.864} &$-0.651\vert1111;0011\rangle + 0.486\vert1111;0101\rangle + 0.442\vert1111;0110\rangle + 0.166\vert1111;1001\rangle + 0.318\vert1111;1010\rangle - 0.124\vert1111;1100\rangle$&1,-1& 0.0\\\cline{3-5}
    &&$-0.461\vert0111;1011\rangle + 0.343\vert0111;1101\rangle + 0.313\vert0111;1110\rangle + 0.461\vert1011;0111\rangle + 0.117\vert1011;1101\rangle + 0.225\vert1011;1110\rangle - 0.343\vert1101;0111\rangle - 0.117\vert1101;1011\rangle - 0.088\vert1101;1110\rangle - 0.313\vert1110;0111\rangle - 0.225\vert1110;1011\rangle + 0.088\vert1110;1101\rangle$&  1,0& 1.0\\\cline{3-5}
    &&$-0.651\vert0011;1111\rangle + 0.486\vert0101;1111\rangle + 0.442\vert0110;1111\rangle + 0.166\vert1001;1111\rangle + 0.318\vert1010;1111\rangle - 0.124\vert1100;1111\rangle$&  1,1& 0.0\\\cline{3-5}
   &&$0.185\vert1111;0011\rangle + 0.472\vert1111;0101\rangle + 0.112\vert1111;0110\rangle - 0.657\vert1111;1001\rangle - 0.327\vert1111;1010\rangle - 0.439\vert1111;1100\rangle$&  1,-1& 0.0\\\cline{3-5}
    &&$-0.131\vert0111;1011\rangle - 0.334\vert0111;1101\rangle - 0.079\vert0111;1110\rangle + 0.131\vert1011;0111\rangle + 0.464\vert1011;1101\rangle + 0.231\vert1011;1110\rangle + 0.334\vert1101;0111\rangle - 0.464\vert1101;1011\rangle + 0.310\vert1101;1110\rangle + 0.079\vert1110;0111\rangle - 0.231\vert1110;1011\rangle - 0.310\vert1110;1101\rangle$&  1,0& 1.0\\\cline{3-5}
     &&$0.185\vert0011;1111\rangle + 0.472\vert01011111\rangle + 0.112\vert0110;1111\rangle - 0.657\vert1001;1111\rangle - 0.327\vert1010;1111\rangle - 0.439\vert1100;1111\rangle$&  1,1& 0.0\\
\end{tabular}
\end{ruledtabular}
\end{table*}

\begin{table*}
  \caption{\label{tab:NV_param_band} Wavefunction parameters, energies (with respect to the ground state triplet, based on PBE DFT calculations), spin moments, and multireference character of the states of \nv{} in the band basis ($\vert \overline{e_xe_ya_1(2)a_1(1)} ; e_xe_ya_1(2)a_1(1)\rangle$).}
  \centering
\begin{ruledtabular}
  \begin{tabular}{c|c|>{\centering\arraybackslash}p{13cm}cc}
   & $\Delta E$ (eV) & Many-body state(s) & $S,m_s$ &$\Lambda_\text{MR}$  \\
    \hline
    \multirow{ 3}{*}{$^3A_2$} & \multirow{ 3}{*}{--} & $\vert1111;0011 \rangle$ & 1,-1& 0.0\\\cline{3-5}
    & & $0.707(\vert0111;1011\rangle - \vert1011;0111\rangle)$ &1,0& 1.0\\\cline{3-5}
    & & $\vert0011;1111\rangle$ &1,1& 0.0\\\hline
    \multirow{ 2}{*}{$^1E$}  &   \multirow{ 2}{*}{0.428} &$-0.498\vert0111;0111\rangle + 0.455\vert0111;1011\rangle - 0.142\vert0111;1101\rangle - 0.066\vert0111;1110\rangle + 0.455\vert1011;0111\rangle + 0.498\vert1011;1011\rangle - 0.130\vert1011;1101\rangle - 0.060\vert1011;1110\rangle - 0.142\vert1101;0111\rangle - 0.130\vert1101;1011\rangle - 0.066\vert1110;0111\rangle - 0.060\vert1110;1011\rangle$ &0,0&1.0\\\cline{3-5}
    &&$-0.455\vert0111;0111\rangle - 0.498\vert0111;1011\rangle - 0.130\vert0111;1101\rangle - 0.060\vert0111;1110\rangle - 0.498\vert1011;0111\rangle + 0.455\vert1011;1011\rangle + 0.142\vert1011;1101\rangle + 0.066\vert1011;1110\rangle - 0.130\vert1101;0111\rangle + 0.142\vert1101;1011\rangle - 0.060\vert1110;0111\rangle + 0.066\vert1110;1011\rangle$ &0,0&1.0\\\hline
    $^1A_1$ & 1.198 & $-0.690\vert0111;0111\rangle - 0.690\vert1011;1011\rangle + 0.158\vert1101;1101\rangle + 0.096\vert1101;1110\rangle + 0.096\vert1110;1101\rangle + 0.061\vert1110;1110\rangle$ & 0,0&1.09\\\hline
    \multirow{ 6}{*}{$^3E$} &\multirow{ 6}{*}{1.864} &$0.681\vert1111;0101\rangle + 0.146\vert1111;0110\rangle - 0.701\vert1111;1001\rangle - 0.150\vert1111;1010\rangle$&1,-1& 0.0\\\cline{3-5}
    &&$-0.482\vert0111;1101\rangle - 0.103\vert0111;1110\rangle + 0.496\vert1011;1101\rangle + 0.106\vert1011;1110\rangle + 0.482\vert1101;0111\rangle - 0.496\vert1101;1011\rangle + 0.103\vert1110;0111\rangle - 0.106\vert1110;1011\rangle$&  1,0& 1.0\\\cline{3-5}
    &&$0.681\vert0101;1111\rangle + 0.146\vert0110;1111\rangle - 0.701\vert1001;1111\rangle - 0.150\vert1010;1111\rangle$&  1,1& 0.0\\\cline{3-5}
    &&$-0.701\vert1111;0101\rangle - 0.150\vert1111;0110\rangle - 0.681\vert1111;1001\rangle - 0.146\vert1111;1010\rangle$&  1,-1& 0.0\\\cline{3-5}
    &&$0.496\vert0111;1101\rangle + 0.106\vert0111;1110\rangle + 0.482\vert1011;1101\rangle + 0.103\vert1011;1110\rangle - 0.496\vert1101;0111\rangle - 0.482\vert1101;1011\rangle - 0.106\vert1110;0111\rangle - 0.103\vert1110;1011\rangle$&  1,0& 1.0\\\cline{3-5}
    &&$-0.701\vert0101;1111\rangle - 0.150\vert01101;111\rangle - 0.681\vert1001;1111\rangle - 0.146\vert1010;1111\rangle$ &  1,1& 0.0\\
\end{tabular}
\end{ruledtabular}
\end{table*}

\subsection{\feal{} in AlN}

As we showed in the main text, the nature of the MB states of \feal{} depend significantly on choice of exchange-correlation functional for the initial DFT calculation, and treatment of the double-counting (DC) correction. For simplicity here, we will just discuss results from PBE calculations with no DC correction added. In any case, the MB wavefunctions for \feal{} in AlN are extensive, and so we do not include them all in Table \ref{tab:feal_param}; instead, we show the state with the maximum $m_s$ for each orbital degeneracy (i.e., $5/2$ for the ground state and $3/2$ for the excited states) for reference, as the high-spin projections are usually the least entangled. Also, the orbital basis of Fe $3d$ projected  (not localized) Wannier functions ($\vert \overline{d_{z^2}}\overline{d_{xz}}\overline{d_{yz}}\overline{d_{x^2-y^2}}\overline{d_{xy}};d_{z^2}d_{xz}d_{yz}d_{x^2-y^2}d_{xy}\rangle$) results in a hopping matrix that is close to diagonal, with only small hoppings between the $d_{xz}$/ $d_{xy}$, and $d_{yz}$ and $d_{x^2-y^2}$ due to the reduced symmetry from $T_d$ to $C_{3v}$. Therefore, there is not a significant qualitative difference between the band and orbital basis and we only report the orbital one.

The only single-determinant states in Table~\ref{tab:feal_param} are the $m_s=\pm5/2$ states of the $^6A_1$ manifold. All other states have $\Lambda_{\text{MR}}>1$. The $m_s=\pm 0.5$ states (not shown) are close to maximally entangled, i.e., $\Lambda^{\text{max}}_{\text{MR}} \simeq 2.5$. 
In addition to the states in the  $^6A_1$ manifold where all spins aligned ($m_s=\pm5/2$), the $m_s=\pm3/2$ states consists of  linear combinations of determinants with one spin flipped, and $m_s=\pm1/2$ consists of linear combinations with three spins flipped. As we see in Table~\ref{tab:feal_param}, the excited states are quite complicated linear combinations even though our basis in this case is somewhat simple.

\begin{table*}
  \caption{\label{tab:feal_param} Wavefunction parameters, energies (with respect to the ground state $^6A_1$, based on PBE DFT calculations), spin moments, and multireference character of the states of \feal{} in wurtzite AlN in the orbital basis ($\vert \overline{d_{z^2}}\overline{d_{xz}}\overline{d_{yz}}\overline{d_{x^2-y^2}}\overline{d_{xy}};d_{z^2}d_{xz}d_{yz}d_{x^2-y^2}d_{xy}\rangle$).}
  \centering
\begin{ruledtabular}
  \begin{tabular}{c|c|>{\centering\arraybackslash}p{13cm}|cc}
   & $\Delta E$ (eV) & Many-body state(s) & $S,m_s$ &$\Lambda_\text{MR}$  \\\hline
    \hline
    $^6A_1$ & -- & $\vert00000;11111\rangle$ & $5/2,5/2$ & 0.0\\\hline
    \multirow{ 2}{*}{$^4E$} & \multirow{ 2}{*}{1.729} & $-0.1\vert01111;00001\rangle + 0.176\vert01111;00010\rangle + 0.148\vert01111;00100\rangle + 0.302\vert01111;01000\rangle + 0.36\vert01111;10000\rangle + 0.12\vert10111;00001\rangle - 0.148\vert10111;00010\rangle + 0.176\vert10111;00100\rangle + 0.36\vert101110;1000\rangle - 0.302\vert10111;10000\rangle - 0.232\vert11011;00001\rangle + 0.023\vert11011;00010\rangle - 0.028\vert11011;00100\rangle - 0.126\vert11011;01000\rangle + 0.106\vert11011;10000\rangle + 0.195\vert11101;00001\rangle - 0.028\vert11101;00010\rangle - 0.023\vert11101;00100\rangle - 0.106\vert11101;01000\rangle - 0.126\vert11101;10000\rangle - 0.312\vert11110;00010\rangle - 0.372\vert11110;00100\rangle - 0.159\vert11110;01000\rangle - 0.133\vert11110;10000\rangle$ & \multirow{ 2}{*}{$3/2,3/2$} & \multirow{ 2}{*}{1.2}\\ \cline{3-3}
    & & $-0.12\vert00001;01111\rangle - 0.1\vert00001;10111\rangle + 0.195\vert00001;11011\rangle + 0.232\vert00001;11101\rangle - 0.148\vert00010;01111\rangle - 0.176\vert00010;10111\rangle + 0.028\vert00010;11011\rangle + 0.023\vert00010;11101\rangle - 0.372\vert00010;11110\rangle + 0.176\vert00100;01111\rangle - 0.148\vert00100;10111\rangle + 0.023\vert00100;11011\rangle - 0.028\vert00100;11101\rangle + 0.312\vert00100;11110\rangle + 0.36\vert01000;01111\rangle - 0.302\vert01000;10111\rangle + 0.106\vert01000;11011\rangle - 0.126\vert01000;11101\rangle + 0.133\vert01000;11110\rangle - 0.302\vert10000;01111\rangle - 0.36\vert10000;10111\rangle + 0.126\vert10000;11011\rangle + 0.106\vert10000;11101\rangle - 0.159\vert10000;11110\rangle$ & & \\\hline
    $^4A_1$ & 1.750 & $ -0.733\vert00001;11110\rangle + 0.263\vert00010;01111\rangle - 0.314\vert00010;11101\rangle - 0.263\vert00100;10111\rangle + 0.314\vert00100;11011\rangle - 0.052\vert01000;10111\rangle + 0.245\vert01000;11011\rangle + 0.052\vert10000;01111\rangle - 0.245\vert10000;11101\rangle$ & $3/2,3/2$ & 1.23\\\hline
    \multirow{ 2}{*}{$^4E$} & \multirow{ 2}{*}{1.898} & $ -0.219\vert00001;01111\rangle + 0.262\vert00001;10111\rangle - 0.164\vert00001;11011\rangle + 0.138\vert00001;11101\rangle - 0.008\vert00010;01111\rangle + 0.006\vert00010;10111\rangle - 0.182\vert00010;11011\rangle + 0.218\vert00010;11101\rangle - 0.268\vert00010;11110\rangle - 0.006\vert00100;01111\rangle - 0.008\vert00100;10111\rangle + 0.218\vert00100;11011\rangle + 0.182\vert00100;11101\rangle - 0.32\vert00100;11110\rangle - 0.248\vert01000;01111\rangle - 0.297\vert01000;10111\rangle + 0.054\vert01000;11011\rangle + 0.045\vert01000;11101\rangle - 0.336\vert01000;11110\rangle - 0.297\vert10000;01111\rangle + 0.248\vert10000;10111\rangle - 0.045\vert10000;11011\rangle + 0.054\vert10000;11101\rangle - 0.281\vert10000;11110\rangle$ & \multirow{ 2}{*}{$3/2,3/2$} & \multirow{ 2}{*}{1.2}\\ \cline{3-3}
    & & $ -0.262\vert0000101111\rangle - 0.219\vert00001;10111\rangle + 0.138\vert00001;11011\rangle + 0.164\vert00001;11101\rangle + 0.006\vert00010;01111\rangle + 0.008\vert00010;10111\rangle - 0.218\vert00010;11011\rangle - 0.182\vert00010;11101\rangle - 0.32\vert00010;11110\rangle - 0.008\vert00100;01111\rangle + 0.006\vert00100;10111\rangle - 0.182\vert00100;11011\rangle + 0.218\vert00100;11101\rangle + 0.268\vert00100;11110\rangle - 0.297\vert01000;01111\rangle + 0.248\vert01000;10111\rangle - 0.045\vert01000;11011\rangle + 0.054\vert01000;11101\rangle + 0.281\vert01000;11110\rangle + 0.248\vert10000;01111\rangle + 0.297\vert10000;10111\rangle - 0.054\vert10000;11011\rangle - 0.045\vert10000;11101\rangle - 0.336\vert10000;11110\rangle$ & & \\\hline
     $^4A_1$ & 1.974 & $0.377\vert00001;11110\rangle + 0.439\vert00010;01111\rangle + 0.257\vert00010;11101\rangle - 0.439\vert00100;10111\rangle - 0.257\vert00100;11011\rangle - 0.068\vert01000;10111\rangle + 0.407\vert01000;11011\rangle + 0.068\vert10000;01111\rangle - 0.407\vert10000;11101\rangle$ & $3/2,3/2$ & 1.53\\\hline
    \multirow{ 2}{*}{$^4E$} & \multirow{ 2}{*}{2.003} & $-0.234\vert00001;01111\rangle + 0.258\vert00001;10111\rangle + 0.151\vert00001;11011\rangle - 0.137\vert00001;11101\rangle - 0.117\vert00010;01111\rangle + 0.106\vert00010;10111\rangle - 0.232\vert00010;11011\rangle + 0.255\vert00010;11101\rangle + 0.329\vert00010;11110\rangle - 0.106\vert00100;01111\rangle - 0.117\vert00100;10111\rangle + 0.255\vert00100;11011\rangle + 0.232\vert00100;11101\rangle + 0.361\vert00100;11110\rangle + 0.202\vert01000;01111\rangle + 0.222\vert01000;10111\rangle + 0.031\vert01000;11011\rangle + 0.028\vert01000;11101\rangle - 0.263\vert01000;11110\rangle + 0.222\vert10000;01111\rangle - 0.202\vert10000;10111\rangle - 0.029\vert10000;11011\rangle + 0.031\vert10000;11101\rangle - 0.239\vert10000;11110\rangle$ & \multirow{ 2}{*}{$3/2,3/2$} & \multirow{ 2}{*}{1.27}\\ \cline{3-3}
    & & $0.258\vert00001;01111\rangle + 0.234\vert00001;10111\rangle + 0.137\vert00001;11011\rangle + 0.151\vert00001;11101\rangle - 0.106\vert00010;01111\rangle - 0.117\vert00010;10111\rangle + 0.255\vert00010;11011\rangle + 0.232\vert00010;11101\rangle - 0.361\vert00010;11110\rangle + 0.117\vert00100;01111\rangle - 0.106\vert00100;10111\rangle + 0.232\vert00100;11011\rangle - 0.255\vert00100;11101\rangle + 0.329\vert00100;11110\rangle - 0.222\vert01000;01111\rangle + 0.202\vert01000;10111\rangle + 0.028\vert01000;11011\rangle - 0.031\vert01000;11101\rangle - 0.239\vert01000;11110\rangle + 0.202\vert10000;01111\rangle + 0.222\vert10000;10111\rangle + 0.031\vert10000;11011\rangle + 0.028\vert10000;11101\rangle + 0.263\vert10000;11110\rangle$ & & \\\hline
$^4A_2$ & 2.022 & $-0.523\vert00010;10111\rangle + 0.006\vert00010;11011\rangle - 0.523\vert00100;01111\rangle + 0.006\vert00100;11101\rangle - 0.015\vert01000;01111\rangle - 0.476\vert01000;11101\rangle - 0.015\vert10000;10111\rangle - 0.476\vert10000;11011\rangle$ & $3/2,3/2$ & 1.50\\
\end{tabular}
\end{ruledtabular}
\end{table*}

\section{Symmetry of many-body states} \label{app:sym}

A crucial property for analyzing the character of the MB states is their symmetry-protected energy degeneracies, which can be identified by determining the irreducible representations of the point group under which the defect states transform. 
We derive the MB representations from the single-particle representations, which can be obtained by applying the symmetry operations to the Wannier functions.
Since we restrict ourselves to systems without SOC, the MB representations split into a tensor product of orbital and spin representations.

Degenerate manifolds of many-body (MB) states calculated by diagonalizing the Hamiltonian in Eq.~(2) of the main text should transform like irreducible representations of the point group of the defect. This is not guaranteed \textit{a priori}, as the Wannierization procedure may break the symmetry. Therefore, we confirm this explicitly  \textit{a posteriori} by calculating the representations of the many-body states resulting from our calculations. The general theory can be found in textbooks (e.g., Ref.~\onlinecite{Avery1976}); here we will give the specific procedure that we use to determine the symmetry of the MB states calculated using the methodology described in Sec.~III of the main text.

Given a symmetry operation $R$ in the symmetry group of the defect, we first determine the single-particle representation of $R$, denoted $D_{ij}(R)$ (where $i$ and $j$ are orbitals in the Wannier basis) by applying the operation to the Wannier functions in real space; since our Wannier function are spinless (i.e., determined from DFT calculations neglecting spin, see Sec.~IV~A of the main text), this will give the orbital part of the representation. $D_{ij}(R)$ can be used to determine the effect of symmetry operations on the creation/annihilation operators, i.e.,
\begin{equation}
  \hat{R}c^\dagger_i\hat{R}^\dagger=\sum_jD_{ij}(R)c^\dagger_j.
  \label{eq:Rsingle}
\end{equation}
We can write a given many-body state as a series of raising operators applied to the vacuum
\begin{equation}
  \vert\Psi_a\rangle = \sum_b w_{ab} c^\dagger_ic^\dagger_jc^\dagger_kc^\dagger_lc^\dagger_m...\vert 0 \rangle
\end{equation}
where $i,j,k,...$ are the occupied spin-orbitals in the Fock state $b$, and $w_{ab}$ is the weight
of the Fock state in the many-body state.
A symmetry operation is applied to the many-body state by applying the single-particle operator $\hat{R}$ to each creation operator
\begin{equation}
  \begin{split}
    \hat{R}\vert\Psi_a\rangle &= \sum_b w_{ab} \hat{R}c^\dagger_i\hat{R}^\dagger\hat{R} c^\dagger_j\hat{R}^\dagger\hat{R} c^\dagger_k\hat{R}^\dagger\hat{R} c^\dagger_l\hat{R}^\dagger\hat{R} c^\dagger_m\hat{R}^\dagger...\hat{R}\vert 0 \rangle\\
    & = \sum_b w_{ab} \hat{R}c^\dagger_i\hat{R}^\dagger\hat{R} c^\dagger_j\hat{R}^\dagger\hat{R} c^\dagger_k\hat{R}^\dagger\hat{R} c^\dagger_l\hat{R}^\dagger\hat{R} c^\dagger_m\hat{R}^\dagger...\vert 0 \rangle
    \end{split}
\end{equation}
where we use that the vacuum is fully symmetric.
Taking the example of \feal{} in AlN (see Sec.~II~C of the main text), which has five electrons, and using  Eq.~(\ref{eq:Rsingle}), we can write this expression as
\begin{equation}
  \begin{split}
    \hat{R}\vert\Psi_a\rangle &= \sum_b w_{ab}
    \sum_{n,o,p,q,r}D_{in}(R)D_{jo}(R)D_{kp}(R)D_{lq}(R)D_{mr}(R)
    c^\dagger_nc^\dagger_oc^\dagger_pc^\dagger_qc^\dagger_r\vert 0 \rangle.
    \end{split}
\end{equation}
Given a degenerate manifold of states, we determine how the states transform by computing the character
\begin{equation}
    \label{chi}
    \chi_{\text{MB}}(R)=\sum_a^{\text{deg.}}\langle\Psi_a\vert\hat{R}\vert\Psi_a\rangle,
\end{equation}
where the sum runs over states in a given degenerate manifold. The characters allow us to decompose the many-body representation into (spinless) irreps of
the point group.

We can include spin in the symmetry analysis by using the fact that a spin-$1/2$ pair transforms under a point-group rotation as
\begin{equation}
\bm{D}_{\text{spin}}(R)= e^{-i\theta\frac{\hat{n}\cdot\bm{\sigma}}{2}},
\end{equation}
where $\theta$ is the angle of rotation, $\bm{\sigma}$ is the vector of Pauli
matrices, and $\hat{n}$ is the normalized axis of the symmetry
operation. Mirror symmetries are the products of inversion and a $\pi$ rotation; since inversion acts trivially on spin, the expression for a mirror symmetry can be obtained by taking $\theta = \pi$ above.
Then, the total representation matrix is
\begin{equation}
  \label{fullrep}
\bm{D}(R)=\bm{D}_{\text{spin}}(R)\otimes\bm{D}_{\text{orb}}(R).
\end{equation}
We can use these spinful representations and the resulting 
symmetry characters to determine the \emph{spinful}
irreps by which degenerate manifolds of states transform. 

\bibliography{defects}